\documentclass[aps,prl,twocolumn,superscriptaddress,amsfont,graphicx,nofootinbib,preprintnumbers]{revtex4-2}

\usepackage{amsmath,amssymb}
\usepackage{graphicx}
\usepackage{xcolor}
\usepackage{slashed}

\begin{document}

\def\Buffalo{Department of Physics, University at Buffalo \\ The State University of New York, Buffalo 14260 USA}

\title{$\mathcal{O}(\alpha\alpha_s^2)$ corrections to the Top Quark mass}

\author{Daniele Gaggero}     
\email[Electronic address:]{dgaggero@buffalo.edu}
\affiliation{\Buffalo}

\author{Ciaran Williams}     
\email[Electronic address:]{ciaranwi@buffalo.edu}
\affiliation{\Buffalo}

\begin{abstract}
We calculate the three-loop $\mathcal{O}(\alpha\alpha_s^2)$ mixed QCD-EW corrections to the top quark mass and obtain the relationships between the 
pole ($M_t$) and $\overline{\rm{MS}}$ ($m_t(\mu)$) masses at this order. In performing our calculation, we express the three-loop self-energies as a function of 
three-loop master integrals (MIs). These MIs are defined so as to solve a differential equation in $\epsilon$-factorized form, allowing for their solution to be expressed as iterated integrals. 
We discuss the types of differential forms that we encounter, which include periods of elliptic curves, and their role in the final physical result. Using our results, we calculate the shift in $m_t(M_t)$ arising at this order. For $M_t = 172.3$ GeV, the shift is -{{216}} MeV, which is comparable to the currently quoted experimental uncertainty. 
\end{abstract}

\maketitle

\section{Introduction}

Thirty years after its discovery~\cite{CDF:1995wbb,D0:1995jca}, the top quark remains a fascinating component of the Standard Model (SM). 
The most notable feature of the top quark is its mass, which at around $172$ GeV is about 35 times heavier than the next heaviest (bottom) quark, and heavier than any other known fundamental particle. 
The origin for this large mass is not currently predicted from first principles and may ultimately tie in with physics Beyond the Standard Model (BSM).  At the LHC the top quark is produced copiously and thus its properties have been measured with increasing precision over the last decade; some recent highlights include studies of its mass~\cite{ATLAS:2024dxp,ATLAS:2025bpp}, width~\cite{ATLAS:2017vgz} and BSM searches~\cite{CMS:2021gfa,CMS:2021hug,ATLAS:2023ujo,CMS:2024ubt,CMS:2022hjj,CMS:2023xyc}.

The heaviness of the top quark has implications for theoretical calculations. More specifically, the lifetime of the top quark is sufficiently small that it decays before it hadronizes. As a result, the top quark mass and width have a much cleaner interpretation as SM parameters than those of the five lighter quarks. Through $\mathcal{O}(\alpha^2)$ the real part of the pole of the fermion propagator can be identified with the pole mass $M_t$, and the imaginary part can be used to extract the width $\Gamma_t$. While the pole mass has less theoretical issues than the equivalent light quarks masses, it is still sensitive to non-perturbative Infrared physics (see Refs~\cite{Beneke:1994sw,Bigi:1994em,Kronfeld:1998di,Beneke:2016cbu}). From the perturbative side, the $\overline{\rm{MS}}$ mass $m_t(\mu)$ provides a somewhat cleaner UV interpretation. The two quantities can be related to one another via perturbation theory, and calculations in this direction have been a mainstay over the past thirty years~\cite{ctx28417338360004803,Gray:1990yh,Fleischer:1998dw,Chetyrkin:1999qi,Melnikov:2000qh,Marquard:2007uj} including  progress in QCD through $\mathcal{O}(\alpha_s^4)$~{\cite{Marquard:2015qpa}, mixed QCD-EW~\cite{Jegerlehner:2003py,Eiras:2005yt},
and two-loop EW corrections~\cite{Kniehl:2014yia}.  The electroweak corrections are particularly interesting; in order to define a gauge invariant top quark mass it is essential to include tadpole diagrams~\cite{Jegerlehner:2003py,Faisst:2004gn}. These terms introduce a large factor of the form $\alpha  N_c  m_t^4/(M_W^2 M_H^2) \sim 0.2$. For physical quantities, like the top Yukawa coupling for instance, these tadpole diagrams also contribute to the renormalization of the vacuum expectation value (vev) and cancel against the similar contributions in the top mass. For this reason, the $\overline{\rm{MS}}$ top mass definition has a slightly unusual theoretical interpretation. When considered in isolation, it has a dramatic dependence on the renormalization scale which comes in part from the significant impact of the tadpole diagrams, but when combined into theoretical observables the issue is significantly abated. In order to side-step these issues at intermediate stages, gauge-dependent tadpole-free mass-schemes have been investigated~\cite{Martin:2016xsp,Martin:2019lqd,Martin:2018emo}. At the cost of being performed in a specific gauge, the resulting parameter has much better perturbative behavior.   
 
In this paper, we present the first calculation of the the 3-loop $\mathcal{O}{(\alpha\alpha_s^2})$ corrections to the quark mass. We work in a Jegerlehner-Kalmykov scheme and fully include the tadpole diagrams, allowing a gauge invariant definition of the $\overline{\rm{MS}}$ mass. Three-loop calculations of EW parameters are complicated by the multiple mass scales in the problem which generate Feynman integrals which include elliptic curves. We discuss the elliptic curves that appear in the MIs at this order and how the periods associated to them can be utilized to generate MIs which satisfy differential equations in $\epsilon$-form.

\section{Calculation}
The top-quark pole mass $M_t$ can be defined by the vanishing of the inverse fermion propagator, namely  
\begin{equation}\label{M_t_def}
    0 = \left[\slashed{q}-m_{0,t}+\Sigma_t(\slashed{q})\right]\big|_{\slashed{q}=M_t} \,,
\end{equation}
where $m_{0,t}$ defines the bare top quark mass and $\Sigma_t(\slashed{q})$ denotes the one-particle irreducible (1PI) diagrams.
For our calculation at $\mathcal{O}(\alpha\alpha_s^2)$, we require the three-loop self-energy, the two-loop self-energies ($\mathcal{O}(\alpha\alpha_s)$ and $\mathcal{O}(\alpha_s^2)$) and their first derivatives with respect to $q^2/m_0^2$, the one-loop $\mathcal{O}(\alpha)$ and $\mathcal{O}(\alpha_s)$ self-energies and their first and second derivatives. The one-loop and two-loop self-energies are well known and we therefore  focus the remainder of our discussion on the more intricate three-loop self-energy. 

\subsection{Three loop Calculation}

We work in the t'Hooft-Feynman gauge, and set $\xi=1$. Feynman diagrams for our calculation can be classified by the type of EW-boson which is exchanged. 
The simplest diagrams correspond to QED corrections, which can be related to the sub-leading color pieces of the $\mathcal{O}(\alpha_s^3)$ calculation and have been studied in the literature~\cite{Chen:2026jid}; accordingly, we will not discuss them further here. The remaining diagrams can be split into two classes, those that involve the exchange of a charged $W$ or $\phi$ boson, and those which involve the exchange of a neutral $Z$, $\chi$ or $H$ boson. For the vast majority of cases, the EW boson couples exclusively to either the external fermion line or an  internal closed loop. Conceptually, these diagrams are straightforward and the majority of the calculation resides in obtaining analytic expressions for the Master Integrals (MIs), which we discuss shortly\footnote{Master integrals with up to five propagators for a generic three-loop quark self-energy have been addressed in \cite{Martin:2021pnd,Martin:2022zlk}.}.
An interesting sub-class of diagrams corresponds to those in which a neutral EW boson connects to the external fermion line and a second internal line. For the chiral couplings associated with the $Z$ or the $\chi$ boson, this presents a challenge since one encounters 
a ${\rm{tr}}(\gamma_5 \gamma^{\mu_1} \dots \gamma^{\mu_n})$ structure, which is not readily defined in $d$-dimensions in a chiral theory. The situation for the $Z$ diagrams is alleviated when one sums over both members of an isodoublet, since the divergent pieces scale like the axial coupling of the individual fermions. Summing over quark doublets thus provides a result which is finite as $\epsilon\rightarrow 0$, and the trace can be taken directly in four-dimensions with no ambiguity.  This leaves the diagrams where the boson is a $\chi$-Goldstone: here, since $m_b=0$, only the top quark contributes and the result contains a divergent piece ($\propto 1/\epsilon)$. 
To avoid an ambiguity in the $\gamma_5$ definition, we construct a counter-term (see~\cite{Freitas:2022hyp}) for each diagram which locally cancels any divergences but does not contain a $\gamma_5$ in its definition. The regrouped term can be evaluated directly in four dimensions, while the integrated counter-term remains in $d$ dimensions, i.e. 
\begin{eqnarray}
\int \prod_{i=1}^{3}{d^d \ell_i}  \frac{\mathcal{N}^{\chi}_{\rm{\gamma_5}}}{\mathcal{D}} \rightarrow&& \bigg(\int \prod_{i=1}^{3}{d^d \ell_i}  \frac{\mathcal{N}^{\chi}_{\rm{\gamma_5}}-\mathcal{N}_{\rm{ct}}}{\mathcal{D}}\bigg)_{\epsilon=0}  \nonumber\\&&+ \int \prod_{i=1}^{3}{d^d \ell_i}  \frac{\mathcal{N}_{\rm{ct}}}{\mathcal{D}} \, ,
\end{eqnarray}
and the trace can safely be taken in four-dimensions.

Each diagram is mapped to a nine propagator auxiliary topology and then reduced using Integration by Parts identities via ${\tt{Kira}}$ \cite{Maierhofer:2017gsa} to a set of MIs. We have created a basis of MIs such that the integrals are in $\epsilon$-form. Specifically, the integrals satisfy the following equation:
\begin{eqnarray}
\label{eq:diff_eps}
d \, \pmb{\mathcal{I}}^{J}_{V}(x,\epsilon) = \epsilon \sum_{i=1}^{j_{\rm{V}}} \mathcal{R}^{J}_i dA^V_{i}  \pmb{\mathcal{I}}^{J}_V(x,\epsilon).
\end{eqnarray}
Here $\pmb{\mathcal{I}}^{J}_{V}(x,\epsilon)$ is an $n_J$-dimensional vector of MIs, associated to either a $V=\phi, W$ or $V=\xi,Z,H$ type diagram ($J$ denotes the topology); $x = M_V^2/m_{t,0}^2$ is a dimensionless variable, $\epsilon$ is the regulating parameter defining excursions away from four-dimensions through $d=4-2\epsilon$, and each $\mathcal{R}^{J}_i$ is a $n_J \times n_J$ rank matrix consisting of rational numbers. Each $dA_i^V$ corresponds to a differential 1-form, with $j_{V}$ denoting the number of unique 1-forms in the topology.   A differential equation\footnote{{Differential equations and iterated integrals have been exploited widely to solve Feynman Integrals, see for example \cite{Kotikov:1990kg,Remiddi:1997ny,GEHRMANN2000485,Laporta:2000dsw,Argeri:2007up,Goncharov:2010jf,Henn:2013pwa,Caron-Huot:2014lda}.}} of the form Eq.~\ref{eq:diff_eps} can be solved iteratively as a series in $\epsilon$, with each term consisting of path integrals from a boundary point $\pmb{\mathcal{I}}^{J}_V(x_0,\epsilon)$:
\begin{eqnarray}
 \pmb{\mathcal{I}}^{J}_{V}(x,\epsilon) &=& \bigg({1} +  \epsilon \sum_{i=1}^{j_{\rm{V}}}  \mathcal{R}^{J}_i  \int_{x_0}^x dA^V_i(t_0)  \nonumber\\&&  +  \epsilon^2 \sum_{i,k=1}^{j_{\rm{V}}}   \mathcal{R}^{J}_k \mathcal{R}^{J}_i  \int_{x_0}^{x} dA^V_k(t_1)  \int_{x_0}^{t_1} dA^V_i(t_0)  \nonumber\\&& + \dots \bigg) \pmb{\mathcal{I}}^{J}_{V}(x_0,\epsilon).
\end{eqnarray}
As such, we observe that the complexity of each MI is set by the individual differential forms $dA_i$ that enter the expression. The simplest forms that we encounter (in the final self-energy expression) are rational polynomials $dx/(x-a)$ where $a \in \{0,\pm 1,2,4 \}$. 
We also encounter the irrational (but well understood) form $dx /\sqrt{x(4\pm x)}$; these roots are readily rationalized by a simple variable change. 
The most complicated differential forms contain periods of elliptic curves, and the nature of the forms is different in the two types of topologies. The set of MIs for several $W$ topologies contains sub-sectors which include the 3-loop on-shell banana integral with one-massless and three massive propagators (of which two internal masses are the top mass). These sub-sectors involve the function $\psi_i(x)$ in the definition of the MIs in $\epsilon$-form. $\psi_{i}(x)$ is a period of an elliptic curve~\cite{Remiddi:2017har,Broedel:2017kkb,Broedel:2018iwv,Weinzierl:2020kyq,Weinzierl:2020fyx,Dlapa:2022wdu}}, which is identical to that found in the calculation of the massive sunrise diagram at two-loops. The resulting differential forms are modular forms. We follow the notation in Ref.~\cite{Honemann:2018mrb} and refer the interested reader to this paper and Refs.~{\cite{Laporta:2004rb,Adams:2013nia,Adams:2015ydq,Remiddi:2016gno,Broedel:2017siw}} for further details. In order to fully define the banana sub-sector for the $W$ topologies, a differential 1-form of modular weight four  is required. However, this differential form does not contribute to the final physical expression. 

A selection of integrals in the $Z$ topologies contains the  3-loop on-shell banana sub-sector, which depends on the period function $\Psi_i(x)$. Following Refs.~\cite{Verrill:1996,Muller-Stach:2012tgj,Pogel:2022yat}, $\Psi_i(x)$ can be written in-terms of $\psi_i$ by identifying
$\Psi_i(y) = \sqrt{y}\psi_i(y) $ and $x = (1 - y) (y-9)/y$. In order to define the basis integrals, two additional functions are needed, which are themselves iterated integrals. Firstly $\mathcal{F}_1$ appears in the banana sub-sector itself 
\begin{eqnarray}
    \mathcal{F}_1(x) &=& \int_{x_0}^x \frac{dz}{\Psi_1^2(z)z\sqrt{64-20z+z^2}} \nonumber\\ && \times \int_0^z  \frac{(w-8)(8+w)^3 \Psi_1^4(w)}{6 \pi^2 (64-20 w + w^2)^2} dw, 
\end{eqnarray}
and is the same function found in the calculation of the equal mass 3-loop banana integral discussed in  Ref~\cite{Pogel:2022yat} (see also~\cite{Broedel:2019kmn,Bonisch:2020qmm,Broedel:2021zij,Pogel:2022vat,Duhr:2025tdf}). Secondly, we need the following function to ensure the vanishing of the $\epsilon^0$ terms of the differential equation for multiple higher propagator MIs: 
\begin{eqnarray}
\mathcal{F}_2(x) = 
   -\int_{x_0}^{x}\frac{z (z+2) \Psi^2_1(z)}{\pi ^2 (z(4 - z))^{3/2}} dz.
\end{eqnarray}
Interestingly, while $\mathcal{F}_1$ is required to define multiple MIs, the final physical expression for the self energy is manifestly independent of this function. Letters containing $\mathcal{F}_2$ do appear in the final expression. The constant $x_0$ is arbitrary and can be chosen to be any value, a natural value for us is $x_0=0$. 

The expansion of the MIs contains multiple iterated integrals of the form
\begin{eqnarray}
I(i_1,\dots i_n;x) = \int_{x_0}^{x} dA^V_{i_1}(t_{n-1}) \,  \dots  \int_{x_0}^{t_1} dA^V_{i_n}(t_0),
\end{eqnarray}
with the complexity of each term being set by the list of letters in the argument. In order to evaluate our $I(i_1,\dots i_n;x)$, which depend on $\psi_1$, $\mathcal{F}_1$ or $\mathcal{F}_2$, we introduce the following parameters:
\begin{eqnarray}
q_V = \exp{(2\pi i\tau_V)}, \; \tau_W = \frac{\psi_{2,0}(x)}{\psi_{1,0}(x)}, \; \tau_Z = \frac{\psi_{2,1}(y)}{6\psi_{1,1}(y)},
\end{eqnarray}
where we have used the notation $\psi_{i,j}(x)$ to denote the $i^{\rm{th}}$ period expanded around the cusp at $x=j$. The periods and variables can be written in terms of Eisenstein series and the Dedekind eta-function, which allows for an all orders representation as a series in the parameter $q_V$. The iterated integral function $I(i_1,\dots i_n;x)$ can thus be evaluated via polynomial integration. We refer the reader to Refs.~\cite{Adams:2013nia,Adams:2015ydq,Remiddi:2016gno,Broedel:2017siw} for further details on this setup. 

Finally, in order to extract the boundary conditions, we have employed several approaches. Two of the simpler methods involve ensuring integrals vanish at appropriate pseudo-thresholds, and, if suitable, develop no imaginary parts ($Z$-type). For the integrals containing differential forms which include the period functions, we have extracted the boundary constants by expanding around a second cusp (e.g. $x=-\infty)$ and matching the solutions for a physical (period independent) integral. This results in a series of simultaneous equations whose solutions yield the boundaries at each cusp. The boundaries at $x=-\infty$ are typically simpler and can be readily fitted using a PSLQ algorithm. This then provides the boundary for the desired integral in terms of known constants and $q_V-$series expansions of the MI at our matching value. We checked our integrals using {\tt{AMFLow}}~\cite{Liu:2022chg}. After validating the two approaches, it was often simpler to use {\tt{AMFLow}} to directly extract high precision inputs for the boundaries, most of which were then readily fitted using a PSLQ algorithm (making extensive use of {\tt{PolyLogTools}}~\cite{Duhr:2019tlz} and {\tt{HandyG}}~\cite{Naterop:2019bzt}).

\section{Renormalization and Validation} 

Solving Eq.~\ref{M_t_def} provides a relationship between the pole mass $M_t$ in terms of the bare mass $m_{0,t}$. In order to remove the UV divergences present in the bare mass and couplings, we perform the following renormalization defining the $\overline{\rm{MS}}{}$ mass $m_t$ and coupling $\alpha_s$. For the latter, 
\begin{eqnarray}
\alpha_{s,0} =  \mu^{2\epsilon} (4\pi)^\epsilon \Gamma(1+\epsilon)\alpha_s(\mu^2) Z_{\alpha_s} .
\end{eqnarray}
Expanding through our desired order, we require 
\begin{eqnarray}
Z_{\alpha_s}
&=& 1
- \frac{1}{\epsilon}\,
\frac{\alpha_s}{4\pi}\beta_0
+\frac{1}{\epsilon}\,
\frac{\alpha_s}{4\pi}
\frac{\alpha}{4\pi}
\bigg(
N_f t_r
\bigg(
\frac{11}{36c_W^2}
+\frac{3}{4s_W^2}\bigg) \nonumber\\&&
-\frac{t_r}{s_W^2}\frac{m_t^2}{m_W^2}
\bigg)+\mathcal{O}(\alpha_s^2) 
\end{eqnarray}
where $\beta_0=\frac{11}{3}C_A-\frac{4}{3}t_r N_f $. To test our setup, we directly calculated the coefficient at $\mathcal{O}(\alpha\alpha_s)$ finding complete agreement with the literature result~\cite{Bednyakov:2012rb}. 
The $\overline{\rm{MS}}$ mass is defined as follows: 
\begin{eqnarray}\label{Zt_def}
    \begin{split}
        m_{t,0} & = m_t (\mu^2) Z_t = m_t(\mu^2) \Bigg(1+\frac{\alpha}{4\pi}\frac{1}{\epsilon}Z_{t,\alpha}^{(1,1)}\\&+\frac{\alpha_s^2}{4\pi}\frac{1}{\epsilon}Z_{t,\alpha_s}^{(1,1)} + \frac{\alpha_s^2}{16\pi^2}\left(\frac{1}{\epsilon^2}Z_{t,\alpha_s^2}^{(2,2)}+\frac{1}{\epsilon}Z_{t,\alpha_s^2}^{(2,1)}\right)\\&  + \frac{\alpha_s}{4\pi}\frac{\alpha}{4\pi}\left(\frac{1}{\epsilon^2}Z_{t,\alpha\alpha_s}^{(2,2)}+\frac{1}{\epsilon}Z_{t,\alpha\alpha_s}^{(2,1)}\right) \\&  + \frac{\alpha_s^2}{16\pi^2}\frac{\alpha}{4\pi}\left(\frac{1}{\epsilon^3}Z_{t,\alpha\alpha_s^2}^{(3,3)}+\frac{1}{\epsilon^2}Z_{t,\alpha\alpha_s^2}^{(3,2)}+\frac{1}{\epsilon}Z_{t,\alpha\alpha_s^2}^{(3,1)}\right)\\&+\mathcal{O}(\alpha^2,\alpha_s^3)\Bigg).
    \end{split}
\end{eqnarray}
Results for the renormalization coefficients for the one and two-loop contributions can be found in Refs.~\cite{ctx28417338360004803,Jegerlehner:2001fb,Jegerlehner:2002em,Jegerlehner:2002er,Gray:1990yh,Fleischer:1998dw,Jegerlehner:2003py}. Using our calculation, we can directly extract $Z^{(3,i)}_{t,\alpha\alpha_s}$. The coefficient of the deepest pole is
\begin{eqnarray}
Z_{t,\alpha\alpha_s^2}^{(3,3)}&=&
\frac{C_F}{16M_H^2M_W^2s_w^2}
\bigg(
2m_t^2
\big(
3(\beta_0+15C_F)M_H^2 \\&&
+16C_A(\beta_0+21C_F)m_t^2
\big)
-(\beta_0+3C_F)P_0
\bigg) \nonumber
\end{eqnarray}
where $P_0$ denotes a polynomial in the EW masses (obtained from the numerator of $Z_{t,\alpha}^{(1,1)}$):
\begin{eqnarray}
P_0 &=&
9M_H^2(M_H^2-m_t^2)+36M_W^4
-24C_A m_t^4  \nonumber\\&&
+8M_H^2(M_Z^2-M_W^2)
+18M_Z^4 \,.
\end{eqnarray}
The renormalization coefficient for the $\epsilon^{-2}$ pole is given by
\begin{align}
    Z_{t,\alpha\alpha_s^2}^{(3,2)}=\frac{C_F\left(9(5\beta_0+14C_A+3C_F)P_0-2P_1\right)}
{864M_h^2M_W^2s_w^2} \,,
\end{align}
where we introduce the polynomial 
\begin{eqnarray}
P_{1} &&= 576 m_t^4 \left(13 C_A^2 + 3 C_A \beta_0 - 6\right) \nonumber\\&&
+ M_h^2 \Big(
27 C_F \left(62 M_W^2 + 19 M_Z^2\right) \nonumber\\&&
+ 11 C_A \left(96 M_W^2 + 66 M_Z^2\right)
\nonumber\\&&- 3 \beta_0 \left(34 M_W^2 + 47 M_Z^2\right) \nonumber\\&&
+ 54 m_t^2 \left(14 C_A + 63 C_F + 9 \beta_0 - 8\right)
\Big).
\end{eqnarray}
Finally, the coefficient of the single pole is found to be
\begin{eqnarray}
\begin{aligned}
Z_{t,\alpha\alpha_s^2}^{(3,1)} &=\frac{C_F}{M_H^2 M_W^2 s_w^2}\Bigg\{\\&
m_t^4\left(
9-8C_A+\frac{5}{2}C_A\beta_0
-\frac{23}{6}C_A C_F
\right)
\\
&+\frac{M_H^2m_t^2}{96}
\left(
200+354C_A+99\beta_0+471C_F
\right)
\\
&+\frac{M_H^2M_W^2}{432}
\left(
4804C_A-879\beta_0+1962C_F
\right)
\\
&+\frac{M_H^2M_Z^2}{864}
\left(
7969C_A-1320\beta_0-7326C_F
\right)
\\
&-\zeta_3\Bigg(
8m_t^4
+\frac{9}{4}m_h^2m_t^2(C_A+2C_F)
\\
&\qquad\qquad
+\frac{m_h^2}{18}
(11C_A-3\beta_0)
(16M_W^2+11M_Z^2)
\Bigg)
\Bigg\}.
\end{aligned}
\end{eqnarray}

We have validated these results for the divergent terms by checking that they match the predictions of the renormalization group equations (RGEs). These equations provide a direct means of calculating the $\epsilon^{-3}$ and $\epsilon^{-2}$ poles from the lower order coefficients {\cite{Jegerlehner:2003py}}. Additionally, the definition of the Yukawa coupling 
provides a means to validate $Z_{t,\alpha\alpha_s^2}^{(3,1)}$; for more details we refer the reader to Refs.~\cite{Jegerlehner:2001fb,Jegerlehner:2002em,Jegerlehner:2002er,Chetyrkin:2012rz,Bednyakov:2012en,Bednyakov:2013eba}.  Finally, we were able to take the imaginary part of our results and use it to extract the NNLO QCD corrections to the top decay width, finding agreement with the literature~\cite{Chen:2022wit,Chen:2023osm}.

\section{Results}
 
After renormalizing our results, we obtain a relationship for the pole mass in terms of the $\overline{\rm{MS}}$ mass. This relationship can be inverted to obtain the (often) more desirable $m_t$ as a function of $M_t$.  We have done so and in this section present results using our derived formula. The complete analytic expression is too long to be effectively reproduced in this format but is attached as an ancillary file to the electronic submission of this manuscript. In addition, we include a Mathematica package which provides a $q_V$ series expansion (through $q_V^{30})$ of the mass and can be utilized for rapid evaluation. 
For our results, we use 
 \begin{eqnarray}
  M_Z &=&  91.188\, {\rm{GeV}},\, \quad   M_W =  80.369, {\rm{GeV}}, \nonumber\\
 M_H &=& 125.13  \, {\rm{GeV}},  \,\quad  \alpha^{-1} = 137.036 .
 \end{eqnarray} 
With these parameters, we find (taking $\mu=M_t$)
 \begin{eqnarray}
 m_t(M_t) \vert_{\mathcal{O}(\alpha\alpha_s^2)}&&=
\alpha\alpha_S^2 C_F\Big(
C_F\left(1.11124-1.18752\,C_A\right) \nonumber\\&& 
-C_A\left(0.250857+1.14576\,C_A\right) \nonumber\\&& 
+N_F\left(0.190193\,C_A-0.0419552\right) 
\Big).
 \end{eqnarray}
   \begin{figure}
  \begin{center}
  \includegraphics[width=8.5cm]{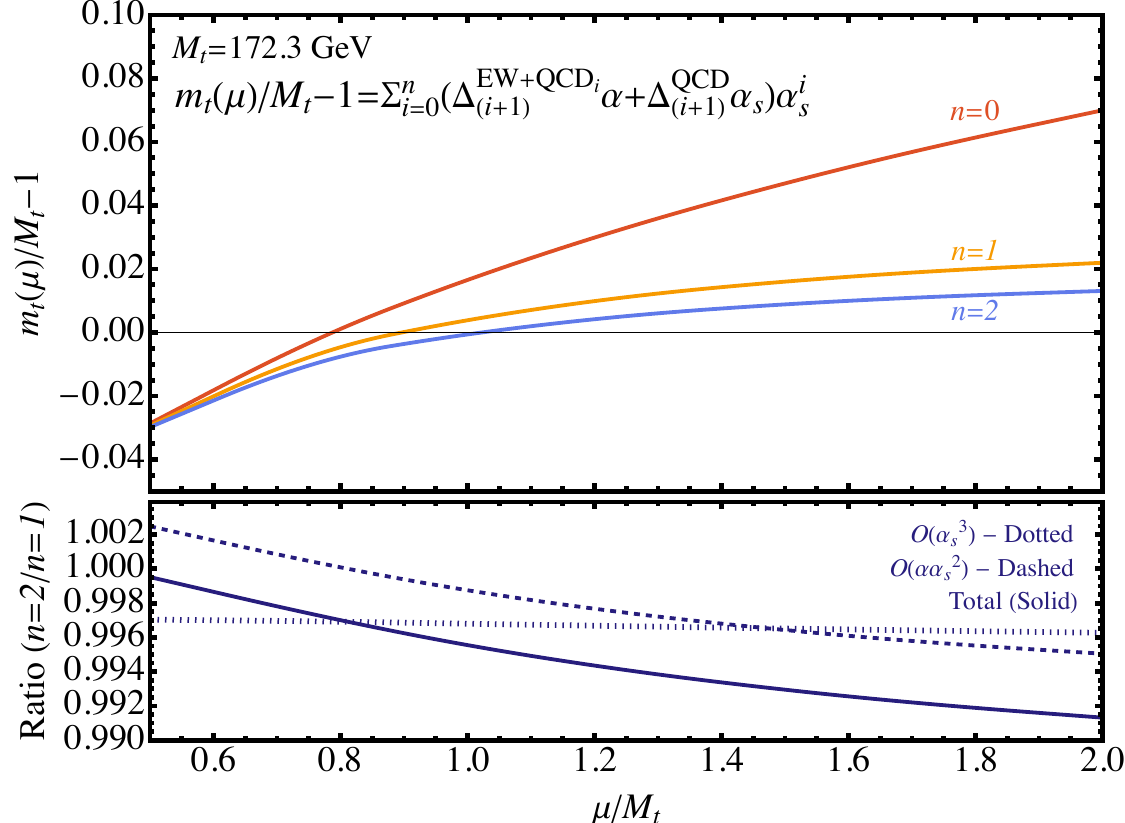}
  \caption{The ratio of the $\overline{\rm{MS}}$ mass $m_t(\mu)$ to the pole mass $M_t$ (minus one), as a function of the renormalization scale. The upper panel displays the corrections grouped as a perturbation series in which the ``lowest" order $(n=0$, red) is defined as the sum of the two one-loop contributions ($\mathcal{O}(\alpha)$ and $\mathcal{O}(\alpha_s)$), as described in the text. When grouped this way, we effectively calculate the NLO $(n=1$, yellow) and NNLO $(n=2$, blue) QCD corrections to this quantity. The lower panel presents the ratio of the $n=2$ to $n=1$ result, taking the QCD (dotted) and mixed-EW (dashed) terms in isolation, as well as the sum of the two (solid). The results for the 3-loop QCD correction, present in the $n=2$ results, are taken from {\tt{RunDec}} 3.0~\cite{Herren:2017dho}.}
  \label{fig:MTresults}
  \end{center}
  \end{figure}
  In Fig.~\ref{fig:MTresults} we present the ratio $m_t(\mu)/M_t-1$ as a function of the renormalization scale $\mu$ in the range $\mu \in \{M_T/2 , 2 M_T\}$. We determine $\alpha_s$ using the 3-loop running and the program {\tt{RunDec}} 3.0~\cite{Herren:2017dho}, taking $\alpha_s(M_Z) = 0.1177$. For simplicity, we have used a 5-flavor running for all scale choices. 
   As is well known~\cite{Jegerlehner:2012kn,Marquard:2016dcn}, there is a large cancellation which happens between the QCD(EW) corrections (which individually reduce(increase) $m_t(\mu)$ for increasing $\mu$). The magnitude and significant contribution to the running mass from the $\mathcal{O}(\alpha)$ corrections, though initially surprising, can be readily understood by considering the role of the tadpole terms. These terms, needed to define a gauge invariant $m_t$, behave like $M_t^4/(M_W^2 M_H^2)$ and generate a logarithmic term which dominates the scale dependence~\cite{PhysRevD.51.1386,Faisst:2004gn,Jegerlehner:2013dpa,Kniehl:2015nwa,Kataev:2022dua}.  Motivated by the large cancelation between the two one-loop contributions, we introduce an effective perturbation series as follows
   \begin{eqnarray}
   m_t(\mu)/M_t -1 = \sum_{i=0}^{\infty} (\Delta^{\rm{EW}+\rm{QCD}}_{(i+1)}\alpha+ \Delta^{\rm{QCD}}_{(i+1)}\alpha_s)\alpha_s^i   \, ,
   \label{eq:pert1}
   \end{eqnarray}
   which is accurate through $\mathcal{O}(\alpha^2)$. We denote the $\overline{\rm{MS}}$ mass obtained from expanding the series up to a fixed $j$ as $m_{t}^{(n=j)}$.  We have defined $\Delta^{\rm{EW}+\rm{QCD}}_j$  as the coefficient of the $\alpha \alpha_s^{j-1}$ contribution to the mass ratio, and $\Delta^{\rm{QCD}}_j$ as the $j^{\rm{th}}$ order pure QCD correction. The primary result of this paper is therefore the calculation of the $\Delta^{\rm{EW}+\rm{QCD}}_3$ contribution, which allows us to expand Eq.~\ref{eq:pert1} through effectively NNLO in QCD.  Our results for this quantity are shown in Fig.~\ref{fig:MTresults} taking $M_t=172.3$ GeV as an input parameter and determining $m_t(\mu)$. At 1-loop ``LO" accuracy there is still a dramatic residual dependence on the renormalization scale, which is dominated by the 1-loop EW contribution; over this range we therefore find that $m^{(n=0)}_t = 175.2^{+9.2}_{-7.9}$ GeV, or around a $\pm 5$\% dependence\footnote{In order to demonstrate the cancellation, this can be compared to the two values if only QCD $m_t^{{\rm{QCD}}_1} = 164.4_{-6.7}^{+8.2}$ GeV and EW  $m_t^{{\rm{EW}}_1} = 183.0_{-16.0}^{+17.0}$ GeV are taken in isolation.}. At 2-loop ``NLO" accuracy we get $m^{(n=1)}_t = 173.0^{+3.1}_{-5.7}$ GeV and, finally, at 3-loops (``NNLO") we obtain $m_t^{(n=2)} = 172.1_{-5.0}^{+2.4}$ GeV.  We find a significant reduction in the scale dependence when expanded through to this order, and perturbative stability. For the central scale choice, we determine
   \begin{eqnarray}
   M_t - m^{(n=2)}_t(M_t) = \rm{106 \; MeV}. 
   \label{eq:mdiffA}
   \end{eqnarray} 
   Shifting from NLO to NNLO accuracy resulted in a shift of  -770 MeV, of which -554 MeV came from the QCD 3-loop corrections and -216 MeV arising from the EW-QCD$^{(2)}$ contribution.  We explore this separation more broadly in the lower panel of Fig.~\ref{fig:MTresults}, where we plot the combined $\Delta^{\rm{EW}+\rm{QCD}}_3$ and $\Delta^{\rm{QCD}}_3$ corrections and each individually, normalized to the $n=1$ result. We observe that our new calculation at $\mathcal{O}(\alpha \alpha_s^2)$ is comparable to the pure QCD $\mathcal{O}(\alpha_s^3)$, and that both correct the ``NLO" result by around 0.5\%.
    
\begin{figure}
  \begin{center}
  \includegraphics[width=8.5cm]{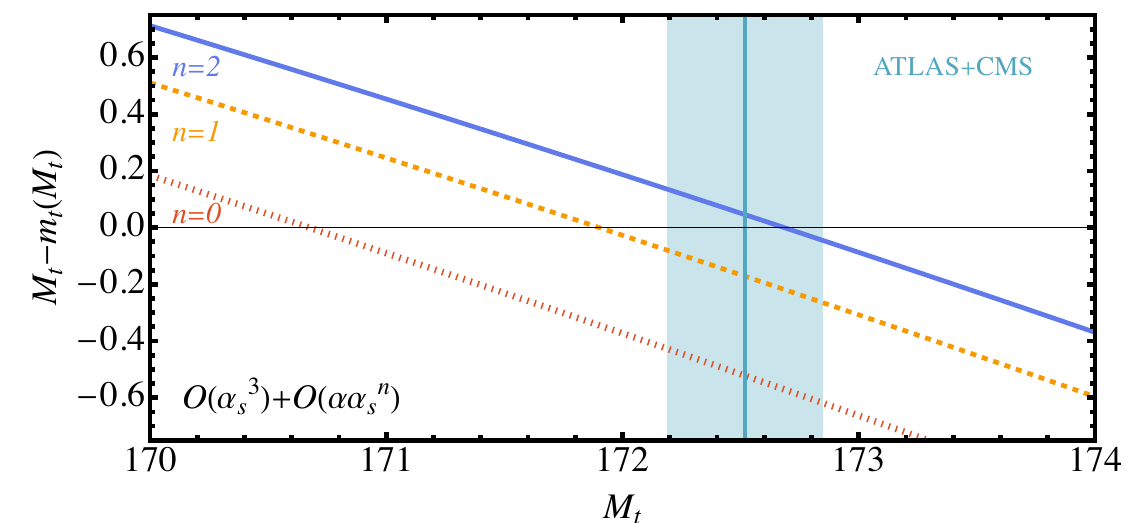}
  \caption{The difference between the Pole mass $M_t$ and the $\overline{\rm{MS}}$ mass $m_t(M_t)$ as a function of the pole mass $M_t$. Each prediction starts from the N3LO QCD prediction and adds in successively higher order $\alpha\alpha_s^n$ corrections. Also shown is the LHC combination measurement of the top quark mass~\cite{ATLAS:2024dxp}.}
  \label{fig:MTmsdiff}
  \end{center}
  \end{figure}

 In Fig.~\ref{fig:MTmsdiff}, we generalize Eq.~\ref{eq:mdiffA} to a wider range of input pole masses; we notice a consistent downward shift of around 200 MeV across the mass range. This is comparable to the experimental accuracy on the recent ATLAS+CMS combination measurement~\cite{ATLAS:2024dxp}
 , which determined $M_t=172.52 \pm 0.33$ GeV. Care should be taken in interpreting this plot however, since the $\mathcal{O}(\alpha^2)$ corrections which are not included in the plot, are known to be sizable~\cite{Jegerlehner:2012kn,Kniehl:2014yia,Kniehl:2015nwa}. 
  
\section{Conclusion} 

We have calculated the 3-loop $\mathcal{O}(\alpha\alpha_s^2)$ correction to the relationship between the pole and $\overline{\rm{MS}}$ top quark masses in the Standard Model. In order to complete our calculation we determined all required MIs as iterated integrals, which arise from the solutions of differential equations in an $\epsilon$-form.  The differential forms which enter the differential equations contain periods associated with the elliptic curves found in the two-loop sunrise ($W$ diagrams) and 3-loop equal mass-banana diagrams $(Z,H$ diagrams). We determined that several differential forms, which are required to fully express the solutions to the differential equations, do not enter the physical prediction. We compared the divergent terms of our result with those predicted by RGEs, finding complete agreement; we also checked that the imaginary part of our result is able to reproduce the top-quark width at NNLO in QCD. We used our calculation to determine $m_t(\mu)$ as a function of $M_t$ (and vice versa) and investigated the impact of our result on the difference between the two masses (with $\mu=M_t$), finding around a 220 MeV shift, which is comparable in size to the current experimental accuracy at the LHC (330 MeV). 

\section*{Acknowledgments}
 
This work was supported by the National Science Foundation through
award NSF-PHY-2310363. Support provided by the Center for Computational Research at the University at Buffalo.

\bibliography{biblio}

\begin{thebibliography}{81}%
\makeatletter
\providecommand \@ifxundefined [1]{%
 \@ifx{#1\undefined}
}%
\providecommand \@ifnum [1]{%
 \ifnum #1\expandafter \@firstoftwo
 \else \expandafter \@secondoftwo
 \fi
}%
\providecommand \@ifx [1]{%
 \ifx #1\expandafter \@firstoftwo
 \else \expandafter \@secondoftwo
 \fi
}%
\providecommand \natexlab [1]{#1}%
\providecommand \enquote  [1]{``#1''}%
\providecommand \bibnamefont  [1]{#1}%
\providecommand \bibfnamefont [1]{#1}%
\providecommand \citenamefont [1]{#1}%
\providecommand \href@noop [0]{\@secondoftwo}%
\providecommand \href [0]{\begingroup \@sanitize@url \@href}%
\providecommand \@href[1]{\@@startlink{#1}\@@href}%
\providecommand \@@href[1]{\endgroup#1\@@endlink}%
\providecommand \@sanitize@url [0]{\catcode `\\12\catcode `\$12\catcode
  `\&12\catcode `\#12\catcode `\^12\catcode `\_12\catcode `\%12\relax}%
\providecommand \@@startlink[1]{}%
\providecommand \@@endlink[0]{}%
\providecommand \url  [0]{\begingroup\@sanitize@url \@url }%
\providecommand \@url [1]{\endgroup\@href {#1}{\urlprefix }}%
\providecommand \urlprefix  [0]{URL }%
\providecommand \Eprint [0]{\href }%
\providecommand \doibase [0]{https://doi.org/}%
\providecommand \selectlanguage [0]{\@gobble}%
\providecommand \bibinfo  [0]{\@secondoftwo}%
\providecommand \bibfield  [0]{\@secondoftwo}%
\providecommand \translation [1]{[#1]}%
\providecommand \BibitemOpen [0]{}%
\providecommand \bibitemStop [0]{}%
\providecommand \bibitemNoStop [0]{.\EOS\space}%
\providecommand \EOS [0]{\spacefactor3000\relax}%
\providecommand \BibitemShut  [1]{\csname bibitem#1\endcsname}%
\let\auto@bib@innerbib\@empty
\bibitem [{\citenamefont {Abe}\ \emph {et~al.}(1995)\citenamefont {Abe} \emph
  {et~al.}}]{CDF:1995wbb}%
  \BibitemOpen
  \bibfield  {author} {\bibinfo {author} {\bibfnamefont {F.}~\bibnamefont
  {Abe}} \emph {et~al.} (\bibinfo {collaboration} {CDF}),\ }\bibfield  {title}
  {\bibinfo {title} {{Observation of top quark production in $\bar{p}p$
  collisions}},\ }\href {https://doi.org/10.1103/PhysRevLett.74.2626}
  {\bibfield  {journal} {\bibinfo  {journal} {Phys. Rev. Lett.}\ }\textbf
  {\bibinfo {volume} {74}},\ \bibinfo {pages} {2626} (\bibinfo {year}
  {1995})},\ \Eprint {https://arxiv.org/abs/hep-ex/9503002}
  {arXiv:hep-ex/9503002} \BibitemShut {NoStop}%
\bibitem [{\citenamefont {Abachi}\ \emph {et~al.}(1995)\citenamefont {Abachi}
  \emph {et~al.}}]{D0:1995jca}%
  \BibitemOpen
  \bibfield  {author} {\bibinfo {author} {\bibfnamefont {S.}~\bibnamefont
  {Abachi}} \emph {et~al.} (\bibinfo {collaboration} {D0}),\ }\bibfield
  {title} {\bibinfo {title} {{Observation of the top quark}},\ }\href
  {https://doi.org/10.1103/PhysRevLett.74.2632} {\bibfield  {journal} {\bibinfo
   {journal} {Phys. Rev. Lett.}\ }\textbf {\bibinfo {volume} {74}},\ \bibinfo
  {pages} {2632} (\bibinfo {year} {1995})},\ \Eprint
  {https://arxiv.org/abs/hep-ex/9503003} {arXiv:hep-ex/9503003} \BibitemShut
  {NoStop}%
\bibitem [{\citenamefont {Hayrapetyan}\ \emph {et~al.}(2024)\citenamefont
  {Hayrapetyan} \emph {et~al.}}]{ATLAS:2024dxp}%
  \BibitemOpen
  \bibfield  {author} {\bibinfo {author} {\bibfnamefont {A.}~\bibnamefont
  {Hayrapetyan}} \emph {et~al.} (\bibinfo {collaboration} {ATLAS, CMS}),\
  }\bibfield  {title} {\bibinfo {title} {{Combination of Measurements of the
  Top Quark Mass from Data Collected by the ATLAS and CMS Experiments at s=7
  and 8~TeV}},\ }\href {https://doi.org/10.1103/PhysRevLett.132.261902}
  {\bibfield  {journal} {\bibinfo  {journal} {Phys. Rev. Lett.}\ }\textbf
  {\bibinfo {volume} {132}},\ \bibinfo {pages} {261902} (\bibinfo {year}
  {2024})},\ \Eprint {https://arxiv.org/abs/2402.08713} {arXiv:2402.08713
  [hep-ex]} \BibitemShut {NoStop}%
\bibitem [{\citenamefont {Aad}\ \emph {et~al.}(2025)\citenamefont {Aad} \emph
  {et~al.}}]{ATLAS:2025bpp}%
  \BibitemOpen
  \bibfield  {author} {\bibinfo {author} {\bibfnamefont {G.}~\bibnamefont
  {Aad}} \emph {et~al.} (\bibinfo {collaboration} {ATLAS}),\ }\bibfield
  {title} {\bibinfo {title} {{Measurement of the top quark mass with the ATLAS
  detector using $t\bar{t}$ events with a high transverse momentum top
  quark}},\ }\href {https://doi.org/10.1016/j.physletb.2025.139608} {\bibfield
  {journal} {\bibinfo  {journal} {Phys. Lett. B}\ }\textbf {\bibinfo {volume}
  {867}},\ \bibinfo {pages} {139608} (\bibinfo {year} {2025})},\ \Eprint
  {https://arxiv.org/abs/2502.18216} {arXiv:2502.18216 [hep-ex]} \BibitemShut
  {NoStop}%
\bibitem [{\citenamefont {Aaboud}\ \emph {et~al.}(2018)\citenamefont {Aaboud}
  \emph {et~al.}}]{ATLAS:2017vgz}%
  \BibitemOpen
  \bibfield  {author} {\bibinfo {author} {\bibfnamefont {M.}~\bibnamefont
  {Aaboud}} \emph {et~al.} (\bibinfo {collaboration} {ATLAS}),\ }\bibfield
  {title} {\bibinfo {title} {{Direct top-quark decay width measurement in the
  $t\bar{t}$ lepton+jets channel at $\sqrt{s}$=8 TeV with the ATLAS
  experiment}},\ }\href {https://doi.org/10.1140/epjc/s10052-018-5595-5}
  {\bibfield  {journal} {\bibinfo  {journal} {Eur. Phys. J. C}\ }\textbf
  {\bibinfo {volume} {78}},\ \bibinfo {pages} {129} (\bibinfo {year} {2018})},\
  \Eprint {https://arxiv.org/abs/1709.04207} {arXiv:1709.04207 [hep-ex]}
  \BibitemShut {NoStop}%
\bibitem [{\citenamefont {Tumasyan}\ \emph
  {et~al.}(2022{\natexlab{a}})\citenamefont {Tumasyan} \emph
  {et~al.}}]{CMS:2021gfa}%
  \BibitemOpen
  \bibfield  {author} {\bibinfo {author} {\bibfnamefont {A.}~\bibnamefont
  {Tumasyan}} \emph {et~al.} (\bibinfo {collaboration} {CMS}),\ }\bibfield
  {title} {\bibinfo {title} {{Search for flavor-changing neutral current
  interactions of the top quark and the Higgs boson decaying to a bottom
  quark-antiquark pair at $ \sqrt{s} $ = 13 TeV}},\ }\href
  {https://doi.org/10.1007/JHEP02(2022)169} {\bibfield  {journal} {\bibinfo
  {journal} {JHEP}\ }\textbf {\bibinfo {volume} {02}},\ \bibinfo {pages}
  {169}},\ \Eprint {https://arxiv.org/abs/2112.09734} {arXiv:2112.09734
  [hep-ex]} \BibitemShut {NoStop}%
\bibitem [{\citenamefont {Tumasyan}\ \emph
  {et~al.}(2022{\natexlab{b}})\citenamefont {Tumasyan} \emph
  {et~al.}}]{CMS:2021hug}%
  \BibitemOpen
  \bibfield  {author} {\bibinfo {author} {\bibfnamefont {A.}~\bibnamefont
  {Tumasyan}} \emph {et~al.} (\bibinfo {collaboration} {CMS}),\ }\bibfield
  {title} {\bibinfo {title} {{Search for Flavor-Changing Neutral Current
  Interactions of the Top Quark and Higgs Boson in Final States with Two
  Photons in Proton-Proton Collisions at $\sqrt{s}=13\text{ }\text{
  }\mathrm{TeV}$}},\ }\href {https://doi.org/10.1103/PhysRevLett.129.032001}
  {\bibfield  {journal} {\bibinfo  {journal} {Phys. Rev. Lett.}\ }\textbf
  {\bibinfo {volume} {129}},\ \bibinfo {pages} {032001} (\bibinfo {year}
  {2022}{\natexlab{b}})},\ \Eprint {https://arxiv.org/abs/2111.02219}
  {arXiv:2111.02219 [hep-ex]} \BibitemShut {NoStop}%
\bibitem [{\citenamefont {Aad}\ \emph {et~al.}(2023)\citenamefont {Aad} \emph
  {et~al.}}]{ATLAS:2023ujo}%
  \BibitemOpen
  \bibfield  {author} {\bibinfo {author} {\bibfnamefont {G.}~\bibnamefont
  {Aad}} \emph {et~al.} (\bibinfo {collaboration} {ATLAS}),\ }\bibfield
  {title} {\bibinfo {title} {{Search for flavour-changing neutral tqH
  interactions with H {\textrightarrow}
  {\ensuremath{\gamma}}{\ensuremath{\gamma}} in pp collisions at $ \sqrt{s} $ =
  13 TeV using the ATLAS detector}},\ }\href
  {https://doi.org/10.1007/JHEP12(2023)195} {\bibfield  {journal} {\bibinfo
  {journal} {JHEP}\ }\textbf {\bibinfo {volume} {12}},\ \bibinfo {pages}
  {195}},\ \Eprint {https://arxiv.org/abs/2309.12817} {arXiv:2309.12817
  [hep-ex]} \BibitemShut {NoStop}%
\bibitem [{\citenamefont {Hayrapetyan}\ \emph {et~al.}(2025)\citenamefont
  {Hayrapetyan} \emph {et~al.}}]{CMS:2024ubt}%
  \BibitemOpen
  \bibfield  {author} {\bibinfo {author} {\bibfnamefont {A.}~\bibnamefont
  {Hayrapetyan}} \emph {et~al.} (\bibinfo {collaboration} {CMS}),\ }\bibfield
  {title} {\bibinfo {title} {{Search for flavor-changing neutral current
  interactions of the top quark mediated by a Higgs boson in proton-proton
  collisions at 13 TeV}},\ }\href {https://doi.org/10.1103/95q6-vvlp}
  {\bibfield  {journal} {\bibinfo  {journal} {Phys. Rev. D}\ }\textbf {\bibinfo
  {volume} {112}},\ \bibinfo {pages} {032008} (\bibinfo {year} {2025})},\
  \Eprint {https://arxiv.org/abs/2407.15172} {arXiv:2407.15172 [hep-ex]}
  \BibitemShut {NoStop}%
\bibitem [{\citenamefont {Tumasyan}\ \emph {et~al.}(2023)\citenamefont
  {Tumasyan} \emph {et~al.}}]{CMS:2022hjj}%
  \BibitemOpen
  \bibfield  {author} {\bibinfo {author} {\bibfnamefont {A.}~\bibnamefont
  {Tumasyan}} \emph {et~al.} (\bibinfo {collaboration} {CMS}),\ }\bibfield
  {title} {\bibinfo {title} {{Search for new physics using effective field
  theory in 13 TeV pp collision events that contain a top quark pair and a
  boosted Z or Higgs boson}},\ }\href
  {https://doi.org/10.1103/PhysRevD.108.032008} {\bibfield  {journal} {\bibinfo
   {journal} {Phys. Rev. D}\ }\textbf {\bibinfo {volume} {108}},\ \bibinfo
  {pages} {032008} (\bibinfo {year} {2023})},\ \Eprint
  {https://arxiv.org/abs/2208.12837} {arXiv:2208.12837 [hep-ex]} \BibitemShut
  {NoStop}%
\bibitem [{\citenamefont {Hayrapetyan}\ \emph {et~al.}(2023)\citenamefont
  {Hayrapetyan} \emph {et~al.}}]{CMS:2023xyc}%
  \BibitemOpen
  \bibfield  {author} {\bibinfo {author} {\bibfnamefont {A.}~\bibnamefont
  {Hayrapetyan}} \emph {et~al.} (\bibinfo {collaboration} {CMS}),\ }\bibfield
  {title} {\bibinfo {title} {{Search for physics beyond the standard model in
  top quark production with additional leptons in the context of effective
  field theory}},\ }\href {https://doi.org/10.1007/JHEP12(2023)068} {\bibfield
  {journal} {\bibinfo  {journal} {JHEP}\ }\textbf {\bibinfo {volume} {12}},\
  \bibinfo {pages} {068}},\ \Eprint {https://arxiv.org/abs/2307.15761}
  {arXiv:2307.15761 [hep-ex]} \BibitemShut {NoStop}%
\bibitem [{\citenamefont {Beneke}\ and\ \citenamefont
  {Braun}(1994)}]{Beneke:1994sw}%
  \BibitemOpen
  \bibfield  {author} {\bibinfo {author} {\bibfnamefont {M.}~\bibnamefont
  {Beneke}}\ and\ \bibinfo {author} {\bibfnamefont {V.~M.}\ \bibnamefont
  {Braun}},\ }\bibfield  {title} {\bibinfo {title} {{Heavy quark effective
  theory beyond perturbation theory: Renormalons, the pole mass and the
  residual mass term}},\ }\href {https://doi.org/10.1016/0550-3213(94)90314-X}
  {\bibfield  {journal} {\bibinfo  {journal} {Nucl. Phys. B}\ }\textbf
  {\bibinfo {volume} {426}},\ \bibinfo {pages} {301} (\bibinfo {year}
  {1994})},\ \Eprint {https://arxiv.org/abs/hep-ph/9402364}
  {arXiv:hep-ph/9402364} \BibitemShut {NoStop}%
\bibitem [{\citenamefont {Bigi}\ \emph {et~al.}(1994)\citenamefont {Bigi},
  \citenamefont {Shifman}, \citenamefont {Uraltsev},\ and\ \citenamefont
  {Vainshtein}}]{Bigi:1994em}%
  \BibitemOpen
  \bibfield  {author} {\bibinfo {author} {\bibfnamefont {I.~I.~Y.}\
  \bibnamefont {Bigi}}, \bibinfo {author} {\bibfnamefont {M.~A.}\ \bibnamefont
  {Shifman}}, \bibinfo {author} {\bibfnamefont {N.~G.}\ \bibnamefont
  {Uraltsev}},\ and\ \bibinfo {author} {\bibfnamefont {A.~I.}\ \bibnamefont
  {Vainshtein}},\ }\bibfield  {title} {\bibinfo {title} {{The Pole mass of the
  heavy quark. Perturbation theory and beyond}},\ }\href
  {https://doi.org/10.1103/PhysRevD.50.2234} {\bibfield  {journal} {\bibinfo
  {journal} {Phys. Rev. D}\ }\textbf {\bibinfo {volume} {50}},\ \bibinfo
  {pages} {2234} (\bibinfo {year} {1994})},\ \Eprint
  {https://arxiv.org/abs/hep-ph/9402360} {arXiv:hep-ph/9402360} \BibitemShut
  {NoStop}%
\bibitem [{\citenamefont {Kronfeld}(1998)}]{Kronfeld:1998di}%
  \BibitemOpen
  \bibfield  {author} {\bibinfo {author} {\bibfnamefont {A.~S.}\ \bibnamefont
  {Kronfeld}},\ }\bibfield  {title} {\bibinfo {title} {{The Perturbative pole
  mass in QCD}},\ }\href {https://doi.org/10.1103/PhysRevD.58.051501}
  {\bibfield  {journal} {\bibinfo  {journal} {Phys. Rev. D}\ }\textbf {\bibinfo
  {volume} {58}},\ \bibinfo {pages} {051501} (\bibinfo {year} {1998})},\
  \Eprint {https://arxiv.org/abs/hep-ph/9805215} {arXiv:hep-ph/9805215}
  \BibitemShut {NoStop}%
\bibitem [{\citenamefont {Beneke}\ \emph {et~al.}(2017)\citenamefont {Beneke},
  \citenamefont {Marquard}, \citenamefont {Nason},\ and\ \citenamefont
  {Steinhauser}}]{Beneke:2016cbu}%
  \BibitemOpen
  \bibfield  {author} {\bibinfo {author} {\bibfnamefont {M.}~\bibnamefont
  {Beneke}}, \bibinfo {author} {\bibfnamefont {P.}~\bibnamefont {Marquard}},
  \bibinfo {author} {\bibfnamefont {P.}~\bibnamefont {Nason}},\ and\ \bibinfo
  {author} {\bibfnamefont {M.}~\bibnamefont {Steinhauser}},\ }\bibfield
  {title} {\bibinfo {title} {{On the ultimate uncertainty of the top quark pole
  mass}},\ }\href {https://doi.org/10.1016/j.physletb.2017.10.054} {\bibfield
  {journal} {\bibinfo  {journal} {Phys. Lett. B}\ }\textbf {\bibinfo {volume}
  {775}},\ \bibinfo {pages} {63} (\bibinfo {year} {2017})},\ \Eprint
  {https://arxiv.org/abs/1605.03609} {arXiv:1605.03609 [hep-ph]} \BibitemShut
  {NoStop}%
\bibitem [{\citenamefont {Tarrach}(81 6)}]{ctx28417338360004803}%
  \BibitemOpen
  \bibfield  {author} {\bibinfo {author} {\bibfnamefont {R.}~\bibnamefont
  {Tarrach}},\ }\bibfield  {title} {\bibinfo {title} {The pole mass in
  perturbative qcd},\ }\href@noop {} {\bibfield  {journal} {\bibinfo  {journal}
  {Nuclear physics.}\ }\textbf {\bibinfo {volume} {183}} (\bibinfo {year}
  {1981-6})}\BibitemShut {NoStop}%
\bibitem [{\citenamefont {Gray}\ \emph {et~al.}(1990)\citenamefont {Gray},
  \citenamefont {Broadhurst}, \citenamefont {Grafe},\ and\ \citenamefont
  {Schilcher}}]{Gray:1990yh}%
  \BibitemOpen
  \bibfield  {author} {\bibinfo {author} {\bibfnamefont {N.}~\bibnamefont
  {Gray}}, \bibinfo {author} {\bibfnamefont {D.~J.}\ \bibnamefont
  {Broadhurst}}, \bibinfo {author} {\bibfnamefont {W.}~\bibnamefont {Grafe}},\
  and\ \bibinfo {author} {\bibfnamefont {K.}~\bibnamefont {Schilcher}},\
  }\bibfield  {title} {\bibinfo {title} {{Three Loop Relation of Quark
  (Modified) Ms and Pole Masses}},\ }\href {https://doi.org/10.1007/BF01614703}
  {\bibfield  {journal} {\bibinfo  {journal} {Z. Phys. C}\ }\textbf {\bibinfo
  {volume} {48}},\ \bibinfo {pages} {673} (\bibinfo {year} {1990})}\BibitemShut
  {NoStop}%
\bibitem [{\citenamefont {Fleischer}\ \emph {et~al.}(1999)\citenamefont
  {Fleischer}, \citenamefont {Jegerlehner}, \citenamefont {Tarasov},\ and\
  \citenamefont {Veretin}}]{Fleischer:1998dw}%
  \BibitemOpen
  \bibfield  {author} {\bibinfo {author} {\bibfnamefont {J.}~\bibnamefont
  {Fleischer}}, \bibinfo {author} {\bibfnamefont {F.}~\bibnamefont
  {Jegerlehner}}, \bibinfo {author} {\bibfnamefont {O.~V.}\ \bibnamefont
  {Tarasov}},\ and\ \bibinfo {author} {\bibfnamefont {O.~L.}\ \bibnamefont
  {Veretin}},\ }\bibfield  {title} {\bibinfo {title} {{Two loop QCD corrections
  of the massive fermion propagator}},\ }\href
  {https://doi.org/10.1016/S0550-3213(98)00705-6} {\bibfield  {journal}
  {\bibinfo  {journal} {Nucl. Phys. B}\ }\textbf {\bibinfo {volume} {539}},\
  \bibinfo {pages} {671} (\bibinfo {year} {1999})},\ \bibinfo {note} {[Erratum:
  Nucl.Phys.B 571, 511--512 (2000)]},\ \Eprint
  {https://arxiv.org/abs/hep-ph/9803493} {arXiv:hep-ph/9803493} \BibitemShut
  {NoStop}%
\bibitem [{\citenamefont {Chetyrkin}\ and\ \citenamefont
  {Steinhauser}(2000)}]{Chetyrkin:1999qi}%
  \BibitemOpen
  \bibfield  {author} {\bibinfo {author} {\bibfnamefont {K.~G.}\ \bibnamefont
  {Chetyrkin}}\ and\ \bibinfo {author} {\bibfnamefont {M.}~\bibnamefont
  {Steinhauser}},\ }\bibfield  {title} {\bibinfo {title} {{The Relation between
  the MS-bar and the on-shell quark mass at order alpha(s)**3}},\ }\href
  {https://doi.org/10.1016/S0550-3213(99)00784-1} {\bibfield  {journal}
  {\bibinfo  {journal} {Nucl. Phys. B}\ }\textbf {\bibinfo {volume} {573}},\
  \bibinfo {pages} {617} (\bibinfo {year} {2000})},\ \Eprint
  {https://arxiv.org/abs/hep-ph/9911434} {arXiv:hep-ph/9911434} \BibitemShut
  {NoStop}%
\bibitem [{\citenamefont {Melnikov}\ and\ \citenamefont
  {Ritbergen}(2000)}]{Melnikov:2000qh}%
  \BibitemOpen
  \bibfield  {author} {\bibinfo {author} {\bibfnamefont {K.}~\bibnamefont
  {Melnikov}}\ and\ \bibinfo {author} {\bibfnamefont {T.~v.}\ \bibnamefont
  {Ritbergen}},\ }\bibfield  {title} {\bibinfo {title} {{The Three loop
  relation between the MS-bar and the pole quark masses}},\ }\href
  {https://doi.org/10.1016/S0370-2693(00)00507-4} {\bibfield  {journal}
  {\bibinfo  {journal} {Phys. Lett. B}\ }\textbf {\bibinfo {volume} {482}},\
  \bibinfo {pages} {99} (\bibinfo {year} {2000})},\ \Eprint
  {https://arxiv.org/abs/hep-ph/9912391} {arXiv:hep-ph/9912391} \BibitemShut
  {NoStop}%
\bibitem [{\citenamefont {Marquard}\ \emph {et~al.}(2007)\citenamefont
  {Marquard}, \citenamefont {Mihaila}, \citenamefont {Piclum},\ and\
  \citenamefont {Steinhauser}}]{Marquard:2007uj}%
  \BibitemOpen
  \bibfield  {author} {\bibinfo {author} {\bibfnamefont {P.}~\bibnamefont
  {Marquard}}, \bibinfo {author} {\bibfnamefont {L.}~\bibnamefont {Mihaila}},
  \bibinfo {author} {\bibfnamefont {J.~H.}\ \bibnamefont {Piclum}},\ and\
  \bibinfo {author} {\bibfnamefont {M.}~\bibnamefont {Steinhauser}},\
  }\bibfield  {title} {\bibinfo {title} {{Relation between the pole and the
  minimally subtracted mass in dimensional regularization and dimensional
  reduction to three-loop order}},\ }\href
  {https://doi.org/10.1016/j.nuclphysb.2007.03.010} {\bibfield  {journal}
  {\bibinfo  {journal} {Nucl. Phys. B}\ }\textbf {\bibinfo {volume} {773}},\
  \bibinfo {pages} {1} (\bibinfo {year} {2007})},\ \Eprint
  {https://arxiv.org/abs/hep-ph/0702185} {arXiv:hep-ph/0702185} \BibitemShut
  {NoStop}%
\bibitem [{\citenamefont {Marquard}\ \emph {et~al.}(2015)\citenamefont
  {Marquard}, \citenamefont {Smirnov}, \citenamefont {Smirnov},\ and\
  \citenamefont {Steinhauser}}]{Marquard:2015qpa}%
  \BibitemOpen
  \bibfield  {author} {\bibinfo {author} {\bibfnamefont {P.}~\bibnamefont
  {Marquard}}, \bibinfo {author} {\bibfnamefont {A.~V.}\ \bibnamefont
  {Smirnov}}, \bibinfo {author} {\bibfnamefont {V.~A.}\ \bibnamefont
  {Smirnov}},\ and\ \bibinfo {author} {\bibfnamefont {M.}~\bibnamefont
  {Steinhauser}},\ }\bibfield  {title} {\bibinfo {title} {{Quark Mass Relations
  to Four-Loop Order in Perturbative QCD}},\ }\href
  {https://doi.org/10.1103/PhysRevLett.114.142002} {\bibfield  {journal}
  {\bibinfo  {journal} {Phys. Rev. Lett.}\ }\textbf {\bibinfo {volume} {114}},\
  \bibinfo {pages} {142002} (\bibinfo {year} {2015})},\ \Eprint
  {https://arxiv.org/abs/1502.01030} {arXiv:1502.01030 [hep-ph]} \BibitemShut
  {NoStop}%
\bibitem [{\citenamefont {Jegerlehner}\ and\ \citenamefont
  {Kalmykov}(2004)}]{Jegerlehner:2003py}%
  \BibitemOpen
  \bibfield  {author} {\bibinfo {author} {\bibfnamefont {F.}~\bibnamefont
  {Jegerlehner}}\ and\ \bibinfo {author} {\bibfnamefont {M.~Y.}\ \bibnamefont
  {Kalmykov}},\ }\bibfield  {title} {\bibinfo {title} {{O(alpha alpha(s))
  correction to the pole mass of the t quark within the standard model}},\
  }\href {https://doi.org/10.1016/j.nuclphysb.2003.10.012} {\bibfield
  {journal} {\bibinfo  {journal} {Nucl. Phys. B}\ }\textbf {\bibinfo {volume}
  {676}},\ \bibinfo {pages} {365} (\bibinfo {year} {2004})},\ \Eprint
  {https://arxiv.org/abs/hep-ph/0308216} {arXiv:hep-ph/0308216} \BibitemShut
  {NoStop}%
\bibitem [{\citenamefont {Eiras}\ and\ \citenamefont
  {Steinhauser}(2006)}]{Eiras:2005yt}%
  \BibitemOpen
  \bibfield  {author} {\bibinfo {author} {\bibfnamefont {D.}~\bibnamefont
  {Eiras}}\ and\ \bibinfo {author} {\bibfnamefont {M.}~\bibnamefont
  {Steinhauser}},\ }\bibfield  {title} {\bibinfo {title} {{Two-loop O(alpha
  alpha(s)) corrections to the on-shell fermion propagator in the standard
  model}},\ }\href {https://doi.org/10.1088/1126-6708/2006/02/010} {\bibfield
  {journal} {\bibinfo  {journal} {JHEP}\ }\textbf {\bibinfo {volume} {02}},\
  \bibinfo {pages} {010}},\ \Eprint {https://arxiv.org/abs/hep-ph/0512099}
  {arXiv:hep-ph/0512099} \BibitemShut {NoStop}%
\bibitem [{\citenamefont {Kniehl}\ and\ \citenamefont
  {Veretin}(2014)}]{Kniehl:2014yia}%
  \BibitemOpen
  \bibfield  {author} {\bibinfo {author} {\bibfnamefont {B.~A.}\ \bibnamefont
  {Kniehl}}\ and\ \bibinfo {author} {\bibfnamefont {O.~L.}\ \bibnamefont
  {Veretin}},\ }\bibfield  {title} {\bibinfo {title} {{Two-loop electroweak
  threshold corrections to the bottom and top Yukawa couplings}},\ }\href
  {https://doi.org/10.1016/j.nuclphysb.2015.02.012} {\bibfield  {journal}
  {\bibinfo  {journal} {Nucl. Phys. B}\ }\textbf {\bibinfo {volume} {885}},\
  \bibinfo {pages} {459} (\bibinfo {year} {2014})},\ \bibinfo {note} {[Erratum:
  Nucl.Phys.B 894, 56--57 (2015)]},\ \Eprint {https://arxiv.org/abs/1401.1844}
  {arXiv:1401.1844 [hep-ph]} \BibitemShut {NoStop}%
\bibitem [{\citenamefont {Faisst}\ \emph {et~al.}(2004)\citenamefont {Faisst},
  \citenamefont {Kuhn},\ and\ \citenamefont {Veretin}}]{Faisst:2004gn}%
  \BibitemOpen
  \bibfield  {author} {\bibinfo {author} {\bibfnamefont {M.}~\bibnamefont
  {Faisst}}, \bibinfo {author} {\bibfnamefont {J.~H.}\ \bibnamefont {Kuhn}},\
  and\ \bibinfo {author} {\bibfnamefont {O.}~\bibnamefont {Veretin}},\
  }\bibfield  {title} {\bibinfo {title} {{Pole versus MS mass definitions in
  the electroweak theory}},\ }\href
  {https://doi.org/10.1016/j.physletb.2004.03.045} {\bibfield  {journal}
  {\bibinfo  {journal} {Phys. Lett. B}\ }\textbf {\bibinfo {volume} {589}},\
  \bibinfo {pages} {35} (\bibinfo {year} {2004})},\ \Eprint
  {https://arxiv.org/abs/hep-ph/0403026} {arXiv:hep-ph/0403026} \BibitemShut
  {NoStop}%
\bibitem [{\citenamefont {Martin}(2016)}]{Martin:2016xsp}%
  \BibitemOpen
  \bibfield  {author} {\bibinfo {author} {\bibfnamefont {S.~P.}\ \bibnamefont
  {Martin}},\ }\bibfield  {title} {\bibinfo {title} {{Top-quark pole mass in
  the tadpole-free $\overline {MS}$ scheme}},\ }\href
  {https://doi.org/10.1103/PhysRevD.93.094017} {\bibfield  {journal} {\bibinfo
  {journal} {Phys. Rev. D}\ }\textbf {\bibinfo {volume} {93}},\ \bibinfo
  {pages} {094017} (\bibinfo {year} {2016})},\ \Eprint
  {https://arxiv.org/abs/1604.01134} {arXiv:1604.01134 [hep-ph]} \BibitemShut
  {NoStop}%
\bibitem [{\citenamefont {Martin}\ and\ \citenamefont
  {Robertson}(2019)}]{Martin:2019lqd}%
  \BibitemOpen
  \bibfield  {author} {\bibinfo {author} {\bibfnamefont {S.~P.}\ \bibnamefont
  {Martin}}\ and\ \bibinfo {author} {\bibfnamefont {D.~G.}\ \bibnamefont
  {Robertson}},\ }\bibfield  {title} {\bibinfo {title} {{Standard model
  parameters in the tadpole-free pure $\overline{\rm{MS}}$ scheme}},\ }\href
  {https://doi.org/10.1103/PhysRevD.100.073004} {\bibfield  {journal} {\bibinfo
   {journal} {Phys. Rev. D}\ }\textbf {\bibinfo {volume} {100}},\ \bibinfo
  {pages} {073004} (\bibinfo {year} {2019})},\ \Eprint
  {https://arxiv.org/abs/1907.02500} {arXiv:1907.02500 [hep-ph]} \BibitemShut
  {NoStop}%
\bibitem [{\citenamefont {Martin}\ and\ \citenamefont
  {Patel}(2018)}]{Martin:2018emo}%
  \BibitemOpen
  \bibfield  {author} {\bibinfo {author} {\bibfnamefont {S.~P.}\ \bibnamefont
  {Martin}}\ and\ \bibinfo {author} {\bibfnamefont {H.~H.}\ \bibnamefont
  {Patel}},\ }\bibfield  {title} {\bibinfo {title} {{Two-loop effective
  potential for generalized gauge fixing}},\ }\href
  {https://doi.org/10.1103/PhysRevD.98.076008} {\bibfield  {journal} {\bibinfo
  {journal} {Phys. Rev. D}\ }\textbf {\bibinfo {volume} {98}},\ \bibinfo
  {pages} {076008} (\bibinfo {year} {2018})},\ \Eprint
  {https://arxiv.org/abs/1808.07615} {arXiv:1808.07615 [hep-ph]} \BibitemShut
  {NoStop}%
\bibitem [{\citenamefont {Chen}\ \emph
  {et~al.}(2026{\natexlab{a}})\citenamefont {Chen}, \citenamefont {Han},
  \citenamefont {Li},\ and\ \citenamefont {Niggetiedt}}]{Chen:2026jid}%
  \BibitemOpen
  \bibfield  {author} {\bibinfo {author} {\bibfnamefont {L.}~\bibnamefont
  {Chen}}, \bibinfo {author} {\bibfnamefont {H.-Y.}\ \bibnamefont {Han}},
  \bibinfo {author} {\bibfnamefont {Z.}~\bibnamefont {Li}},\ and\ \bibinfo
  {author} {\bibfnamefont {M.}~\bibnamefont {Niggetiedt}},\ }\bibfield  {title}
  {\bibinfo {title} {{Three-loop QCD+QED corrections to on-shell quark
  renormalization}},\ }\href {https://doi.org/10.1007/JHEP09(2026)126}
  {\bibfield  {journal} {\bibinfo  {journal} {JHEP}\ }\textbf {\bibinfo
  {volume} {09}},\ \bibinfo {pages} {126}},\ \Eprint
  {https://arxiv.org/abs/2602.10973} {arXiv:2602.10973 [hep-ph]} \BibitemShut
  {NoStop}%
\bibitem [{\citenamefont {Martin}(2022)}]{Martin:2021pnd}%
  \BibitemOpen
  \bibfield  {author} {\bibinfo {author} {\bibfnamefont {S.~P.}\ \bibnamefont
  {Martin}},\ }\bibfield  {title} {\bibinfo {title} {{Renormalized
  {\ensuremath{\varepsilon}}-finite master integrals and their virtues: The
  three-loop self-energy case}},\ }\href
  {https://doi.org/10.1103/PhysRevD.105.056014} {\bibfield  {journal} {\bibinfo
   {journal} {Phys. Rev. D}\ }\textbf {\bibinfo {volume} {105}},\ \bibinfo
  {pages} {056014} (\bibinfo {year} {2022})},\ \Eprint
  {https://arxiv.org/abs/2112.07694} {arXiv:2112.07694 [hep-ph]} \BibitemShut
  {NoStop}%
\bibitem [{\citenamefont {Martin}(2023)}]{Martin:2022zlk}%
  \BibitemOpen
  \bibfield  {author} {\bibinfo {author} {\bibfnamefont {S.~P.}\ \bibnamefont
  {Martin}},\ }\bibfield  {title} {\bibinfo {title} {{Evaluation of three-loop
  self-energy master integrals with four or five propagators}},\ }\href
  {https://doi.org/10.1103/PhysRevD.107.053005} {\bibfield  {journal} {\bibinfo
   {journal} {Phys. Rev. D}\ }\textbf {\bibinfo {volume} {107}},\ \bibinfo
  {pages} {053005} (\bibinfo {year} {2023})},\ \Eprint
  {https://arxiv.org/abs/2211.16539} {arXiv:2211.16539 [hep-ph]} \BibitemShut
  {NoStop}%
\bibitem [{\citenamefont {Freitas}\ and\ \citenamefont
  {Song}(2023)}]{Freitas:2022hyp}%
  \BibitemOpen
  \bibfield  {author} {\bibinfo {author} {\bibfnamefont {A.}~\bibnamefont
  {Freitas}}\ and\ \bibinfo {author} {\bibfnamefont {Q.}~\bibnamefont {Song}},\
  }\bibfield  {title} {\bibinfo {title} {{Two-Loop Electroweak Corrections with
  Fermion Loops to e+e-{\textrightarrow}ZH}},\ }\href
  {https://doi.org/10.1103/PhysRevLett.130.031801} {\bibfield  {journal}
  {\bibinfo  {journal} {Phys. Rev. Lett.}\ }\textbf {\bibinfo {volume} {130}},\
  \bibinfo {pages} {031801} (\bibinfo {year} {2023})},\ \Eprint
  {https://arxiv.org/abs/2209.07612} {arXiv:2209.07612 [hep-ph]} \BibitemShut
  {NoStop}%
\bibitem [{\citenamefont {Maierh\"ofer}\ \emph {et~al.}(2018)\citenamefont
  {Maierh\"ofer}, \citenamefont {Usovitsch},\ and\ \citenamefont
  {Uwer}}]{Maierhofer:2017gsa}%
  \BibitemOpen
  \bibfield  {author} {\bibinfo {author} {\bibfnamefont {P.}~\bibnamefont
  {Maierh\"ofer}}, \bibinfo {author} {\bibfnamefont {J.}~\bibnamefont
  {Usovitsch}},\ and\ \bibinfo {author} {\bibfnamefont {P.}~\bibnamefont
  {Uwer}},\ }\bibfield  {title} {\bibinfo {title} {{Kira\textemdash{}A Feynman
  integral reduction program}},\ }\href
  {https://doi.org/10.1016/j.cpc.2018.04.012} {\bibfield  {journal} {\bibinfo
  {journal} {Comput. Phys. Commun.}\ }\textbf {\bibinfo {volume} {230}},\
  \bibinfo {pages} {99} (\bibinfo {year} {2018})},\ \Eprint
  {https://arxiv.org/abs/1705.05610} {arXiv:1705.05610 [hep-ph]} \BibitemShut
  {NoStop}%
\bibitem [{\citenamefont {Kotikov}(1991)}]{Kotikov:1990kg}%
  \BibitemOpen
  \bibfield  {author} {\bibinfo {author} {\bibfnamefont {A.~V.}\ \bibnamefont
  {Kotikov}},\ }\bibfield  {title} {\bibinfo {title} {{Differential equations
  method: New technique for massive Feynman diagrams calculation}},\ }\href
  {https://doi.org/10.1016/0370-2693(91)90413-K} {\bibfield  {journal}
  {\bibinfo  {journal} {Phys. Lett. B}\ }\textbf {\bibinfo {volume} {254}},\
  \bibinfo {pages} {158} (\bibinfo {year} {1991})}\BibitemShut {NoStop}%
\bibitem [{\citenamefont {Remiddi}(1997)}]{Remiddi:1997ny}%
  \BibitemOpen
  \bibfield  {author} {\bibinfo {author} {\bibfnamefont {E.}~\bibnamefont
  {Remiddi}},\ }\bibfield  {title} {\bibinfo {title} {{Differential equations
  for Feynman graph amplitudes}},\ }\href {https://doi.org/10.1007/BF03185566}
  {\bibfield  {journal} {\bibinfo  {journal} {Nuovo Cim. A}\ }\textbf {\bibinfo
  {volume} {110}},\ \bibinfo {pages} {1435} (\bibinfo {year} {1997})},\ \Eprint
  {https://arxiv.org/abs/hep-th/9711188} {arXiv:hep-th/9711188} \BibitemShut
  {NoStop}%
\bibitem [{\citenamefont {Gehrmann}\ and\ \citenamefont
  {Remiddi}(2000)}]{GEHRMANN2000485}%
  \BibitemOpen
  \bibfield  {author} {\bibinfo {author} {\bibfnamefont {T.}~\bibnamefont
  {Gehrmann}}\ and\ \bibinfo {author} {\bibfnamefont {E.}~\bibnamefont
  {Remiddi}},\ }\bibfield  {title} {\bibinfo {title} {Differential equations
  for two-loop four-point functions},\ }\href
  {https://doi.org/https://doi.org/10.1016/S0550-3213(00)00223-6} {\bibfield
  {journal} {\bibinfo  {journal} {Nuclear Physics B}\ }\textbf {\bibinfo
  {volume} {580}},\ \bibinfo {pages} {485} (\bibinfo {year}
  {2000})}\BibitemShut {NoStop}%
\bibitem [{\citenamefont {Laporta}(2000)}]{Laporta:2000dsw}%
  \BibitemOpen
  \bibfield  {author} {\bibinfo {author} {\bibfnamefont {S.}~\bibnamefont
  {Laporta}},\ }\bibfield  {title} {\bibinfo {title} {{High precision
  calculation of multiloop Feynman integrals by difference equations}},\ }\href
  {https://doi.org/10.1142/S0217751X00002159} {\bibfield  {journal} {\bibinfo
  {journal} {Int. J. Mod. Phys. A}\ }\textbf {\bibinfo {volume} {15}},\
  \bibinfo {pages} {5087} (\bibinfo {year} {2000})},\ \Eprint
  {https://arxiv.org/abs/hep-ph/0102033} {arXiv:hep-ph/0102033} \BibitemShut
  {NoStop}%
\bibitem [{\citenamefont {Argeri}\ and\ \citenamefont
  {Mastrolia}(2007)}]{Argeri:2007up}%
  \BibitemOpen
  \bibfield  {author} {\bibinfo {author} {\bibfnamefont {M.}~\bibnamefont
  {Argeri}}\ and\ \bibinfo {author} {\bibfnamefont {P.}~\bibnamefont
  {Mastrolia}},\ }\bibfield  {title} {\bibinfo {title} {{Feynman Diagrams and
  Differential Equations}},\ }\href {https://doi.org/10.1142/S0217751X07037147}
  {\bibfield  {journal} {\bibinfo  {journal} {Int. J. Mod. Phys. A}\ }\textbf
  {\bibinfo {volume} {22}},\ \bibinfo {pages} {4375} (\bibinfo {year}
  {2007})},\ \Eprint {https://arxiv.org/abs/0707.4037} {arXiv:0707.4037
  [hep-ph]} \BibitemShut {NoStop}%
\bibitem [{\citenamefont {Goncharov}\ \emph {et~al.}(2010)\citenamefont
  {Goncharov}, \citenamefont {Spradlin}, \citenamefont {Vergu},\ and\
  \citenamefont {Volovich}}]{Goncharov:2010jf}%
  \BibitemOpen
  \bibfield  {author} {\bibinfo {author} {\bibfnamefont {A.~B.}\ \bibnamefont
  {Goncharov}}, \bibinfo {author} {\bibfnamefont {M.}~\bibnamefont {Spradlin}},
  \bibinfo {author} {\bibfnamefont {C.}~\bibnamefont {Vergu}},\ and\ \bibinfo
  {author} {\bibfnamefont {A.}~\bibnamefont {Volovich}},\ }\bibfield  {title}
  {\bibinfo {title} {{Classical Polylogarithms for Amplitudes and Wilson
  Loops}},\ }\href {https://doi.org/10.1103/PhysRevLett.105.151605} {\bibfield
  {journal} {\bibinfo  {journal} {Phys. Rev. Lett.}\ }\textbf {\bibinfo
  {volume} {105}},\ \bibinfo {pages} {151605} (\bibinfo {year} {2010})},\
  \Eprint {https://arxiv.org/abs/1006.5703} {arXiv:1006.5703 [hep-th]}
  \BibitemShut {NoStop}%
\bibitem [{\citenamefont {Henn}(2013)}]{Henn:2013pwa}%
  \BibitemOpen
  \bibfield  {author} {\bibinfo {author} {\bibfnamefont {J.~M.}\ \bibnamefont
  {Henn}},\ }\bibfield  {title} {\bibinfo {title} {{Multiloop integrals in
  dimensional regularization made simple}},\ }\href
  {https://doi.org/10.1103/PhysRevLett.110.251601} {\bibfield  {journal}
  {\bibinfo  {journal} {Phys. Rev. Lett.}\ }\textbf {\bibinfo {volume} {110}},\
  \bibinfo {pages} {251601} (\bibinfo {year} {2013})},\ \Eprint
  {https://arxiv.org/abs/1304.1806} {arXiv:1304.1806 [hep-th]} \BibitemShut
  {NoStop}%
\bibitem [{\citenamefont {Caron-Huot}\ and\ \citenamefont
  {Henn}(2014)}]{Caron-Huot:2014lda}%
  \BibitemOpen
  \bibfield  {author} {\bibinfo {author} {\bibfnamefont {S.}~\bibnamefont
  {Caron-Huot}}\ and\ \bibinfo {author} {\bibfnamefont {J.~M.}\ \bibnamefont
  {Henn}},\ }\bibfield  {title} {\bibinfo {title} {{Iterative structure of
  finite loop integrals}},\ }\href {https://doi.org/10.1007/JHEP06(2014)114}
  {\bibfield  {journal} {\bibinfo  {journal} {JHEP}\ }\textbf {\bibinfo
  {volume} {06}},\ \bibinfo {pages} {114}},\ \Eprint
  {https://arxiv.org/abs/1404.2922} {arXiv:1404.2922 [hep-th]} \BibitemShut
  {NoStop}%
\bibitem [{\citenamefont {Remiddi}\ and\ \citenamefont
  {Tancredi}(2017)}]{Remiddi:2017har}%
  \BibitemOpen
  \bibfield  {author} {\bibinfo {author} {\bibfnamefont {E.}~\bibnamefont
  {Remiddi}}\ and\ \bibinfo {author} {\bibfnamefont {L.}~\bibnamefont
  {Tancredi}},\ }\bibfield  {title} {\bibinfo {title} {{An Elliptic
  Generalization of Multiple Polylogarithms}},\ }\href
  {https://doi.org/10.1016/j.nuclphysb.2017.10.007} {\bibfield  {journal}
  {\bibinfo  {journal} {Nucl. Phys. B}\ }\textbf {\bibinfo {volume} {925}},\
  \bibinfo {pages} {212} (\bibinfo {year} {2017})},\ \Eprint
  {https://arxiv.org/abs/1709.03622} {arXiv:1709.03622 [hep-ph]} \BibitemShut
  {NoStop}%
\bibitem [{\citenamefont {Broedel}\ \emph
  {et~al.}(2018{\natexlab{a}})\citenamefont {Broedel}, \citenamefont {Duhr},
  \citenamefont {Dulat},\ and\ \citenamefont {Tancredi}}]{Broedel:2017kkb}%
  \BibitemOpen
  \bibfield  {author} {\bibinfo {author} {\bibfnamefont {J.}~\bibnamefont
  {Broedel}}, \bibinfo {author} {\bibfnamefont {C.}~\bibnamefont {Duhr}},
  \bibinfo {author} {\bibfnamefont {F.}~\bibnamefont {Dulat}},\ and\ \bibinfo
  {author} {\bibfnamefont {L.}~\bibnamefont {Tancredi}},\ }\bibfield  {title}
  {\bibinfo {title} {{Elliptic polylogarithms and iterated integrals on
  elliptic curves. Part I: general formalism}},\ }\href
  {https://doi.org/10.1007/JHEP05(2018)093} {\bibfield  {journal} {\bibinfo
  {journal} {JHEP}\ }\textbf {\bibinfo {volume} {05}},\ \bibinfo {pages}
  {093}},\ \Eprint {https://arxiv.org/abs/1712.07089} {arXiv:1712.07089
  [hep-th]} \BibitemShut {NoStop}%
\bibitem [{\citenamefont {Broedel}\ \emph
  {et~al.}(2018{\natexlab{b}})\citenamefont {Broedel}, \citenamefont {Duhr},
  \citenamefont {Dulat}, \citenamefont {Penante},\ and\ \citenamefont
  {Tancredi}}]{Broedel:2018iwv}%
  \BibitemOpen
  \bibfield  {author} {\bibinfo {author} {\bibfnamefont {J.}~\bibnamefont
  {Broedel}}, \bibinfo {author} {\bibfnamefont {C.}~\bibnamefont {Duhr}},
  \bibinfo {author} {\bibfnamefont {F.}~\bibnamefont {Dulat}}, \bibinfo
  {author} {\bibfnamefont {B.}~\bibnamefont {Penante}},\ and\ \bibinfo {author}
  {\bibfnamefont {L.}~\bibnamefont {Tancredi}},\ }\bibfield  {title} {\bibinfo
  {title} {{Elliptic symbol calculus: from elliptic polylogarithms to iterated
  integrals of Eisenstein series}},\ }\href
  {https://doi.org/10.1007/JHEP08(2018)014} {\bibfield  {journal} {\bibinfo
  {journal} {JHEP}\ }\textbf {\bibinfo {volume} {08}},\ \bibinfo {pages}
  {014}},\ \Eprint {https://arxiv.org/abs/1803.10256} {arXiv:1803.10256
  [hep-th]} \BibitemShut {NoStop}%
\bibitem [{\citenamefont {Weinzierl}(2020)}]{Weinzierl:2020kyq}%
  \BibitemOpen
  \bibfield  {author} {\bibinfo {author} {\bibfnamefont {S.}~\bibnamefont
  {Weinzierl}},\ }\bibfield  {title} {\bibinfo {title} {{Iterated Integrals
  Related to Feynman Integrals Associated to Elliptic Curves}},\ }in\ \href
  {https://doi.org/10.1007/978-3-030-80219-6_20} {\emph {\bibinfo {booktitle}
  {{Antidifferentiation and the Calculation of Feynman Amplitudes}}}}\
  (\bibinfo {year} {2020})\ \Eprint {https://arxiv.org/abs/2012.08429}
  {arXiv:2012.08429 [hep-th]} \BibitemShut {NoStop}%
\bibitem [{\citenamefont {Weinzierl}(2021)}]{Weinzierl:2020fyx}%
  \BibitemOpen
  \bibfield  {author} {\bibinfo {author} {\bibfnamefont {S.}~\bibnamefont
  {Weinzierl}},\ }\bibfield  {title} {\bibinfo {title} {{Modular
  transformations of elliptic Feynman integrals}},\ }\href
  {https://doi.org/10.1016/j.nuclphysb.2021.115309} {\bibfield  {journal}
  {\bibinfo  {journal} {Nucl. Phys. B}\ }\textbf {\bibinfo {volume} {964}},\
  \bibinfo {pages} {115309} (\bibinfo {year} {2021})},\ \Eprint
  {https://arxiv.org/abs/2011.07311} {arXiv:2011.07311 [hep-th]} \BibitemShut
  {NoStop}%
\bibitem [{\citenamefont {Dlapa}\ \emph {et~al.}(2023)\citenamefont {Dlapa},
  \citenamefont {Henn},\ and\ \citenamefont {Wagner}}]{Dlapa:2022wdu}%
  \BibitemOpen
  \bibfield  {author} {\bibinfo {author} {\bibfnamefont {C.}~\bibnamefont
  {Dlapa}}, \bibinfo {author} {\bibfnamefont {J.~M.}\ \bibnamefont {Henn}},\
  and\ \bibinfo {author} {\bibfnamefont {F.~J.}\ \bibnamefont {Wagner}},\
  }\bibfield  {title} {\bibinfo {title} {{An algorithmic approach to finding
  canonical differential equations for elliptic Feynman integrals}},\ }\href
  {https://doi.org/10.1007/JHEP08(2023)120} {\bibfield  {journal} {\bibinfo
  {journal} {JHEP}\ }\textbf {\bibinfo {volume} {08}},\ \bibinfo {pages}
  {120}},\ \Eprint {https://arxiv.org/abs/2211.16357} {arXiv:2211.16357
  [hep-ph]} \BibitemShut {NoStop}%
\bibitem [{\citenamefont {H\"onemann}\ \emph {et~al.}(2018)\citenamefont
  {H\"onemann}, \citenamefont {Tempest},\ and\ \citenamefont
  {Weinzierl}}]{Honemann:2018mrb}%
  \BibitemOpen
  \bibfield  {author} {\bibinfo {author} {\bibfnamefont {I.}~\bibnamefont
  {H\"onemann}}, \bibinfo {author} {\bibfnamefont {K.}~\bibnamefont
  {Tempest}},\ and\ \bibinfo {author} {\bibfnamefont {S.}~\bibnamefont
  {Weinzierl}},\ }\bibfield  {title} {\bibinfo {title} {{Electron self-energy
  in QED at two loops revisited}},\ }\href
  {https://doi.org/10.1103/PhysRevD.98.113008} {\bibfield  {journal} {\bibinfo
  {journal} {Phys. Rev. D}\ }\textbf {\bibinfo {volume} {98}},\ \bibinfo
  {pages} {113008} (\bibinfo {year} {2018})},\ \Eprint
  {https://arxiv.org/abs/1811.09308} {arXiv:1811.09308 [hep-ph]} \BibitemShut
  {NoStop}%
\bibitem [{\citenamefont {Laporta}\ and\ \citenamefont
  {Remiddi}(2005)}]{Laporta:2004rb}%
  \BibitemOpen
  \bibfield  {author} {\bibinfo {author} {\bibfnamefont {S.}~\bibnamefont
  {Laporta}}\ and\ \bibinfo {author} {\bibfnamefont {E.}~\bibnamefont
  {Remiddi}},\ }\bibfield  {title} {\bibinfo {title} {{Analytic treatment of
  the two loop equal mass sunrise graph}},\ }\href
  {https://doi.org/10.1016/j.nuclphysb.2004.10.044} {\bibfield  {journal}
  {\bibinfo  {journal} {Nucl. Phys. B}\ }\textbf {\bibinfo {volume} {704}},\
  \bibinfo {pages} {349} (\bibinfo {year} {2005})},\ \Eprint
  {https://arxiv.org/abs/hep-ph/0406160} {arXiv:hep-ph/0406160} \BibitemShut
  {NoStop}%
\bibitem [{\citenamefont {Adams}\ \emph {et~al.}(2013)\citenamefont {Adams},
  \citenamefont {Bogner},\ and\ \citenamefont {Weinzierl}}]{Adams:2013nia}%
  \BibitemOpen
  \bibfield  {author} {\bibinfo {author} {\bibfnamefont {L.}~\bibnamefont
  {Adams}}, \bibinfo {author} {\bibfnamefont {C.}~\bibnamefont {Bogner}},\ and\
  \bibinfo {author} {\bibfnamefont {S.}~\bibnamefont {Weinzierl}},\ }\bibfield
  {title} {\bibinfo {title} {{The two-loop sunrise graph with arbitrary
  masses}},\ }\href {https://doi.org/10.1063/1.4804996} {\bibfield  {journal}
  {\bibinfo  {journal} {J. Math. Phys.}\ }\textbf {\bibinfo {volume} {54}},\
  \bibinfo {pages} {052303} (\bibinfo {year} {2013})},\ \Eprint
  {https://arxiv.org/abs/1302.7004} {arXiv:1302.7004 [hep-ph]} \BibitemShut
  {NoStop}%
\bibitem [{\citenamefont {Adams}\ \emph {et~al.}(2016)\citenamefont {Adams},
  \citenamefont {Bogner},\ and\ \citenamefont {Weinzierl}}]{Adams:2015ydq}%
  \BibitemOpen
  \bibfield  {author} {\bibinfo {author} {\bibfnamefont {L.}~\bibnamefont
  {Adams}}, \bibinfo {author} {\bibfnamefont {C.}~\bibnamefont {Bogner}},\ and\
  \bibinfo {author} {\bibfnamefont {S.}~\bibnamefont {Weinzierl}},\ }\bibfield
  {title} {\bibinfo {title} {{The iterated structure of the all-order result
  for the two-loop sunrise integral}},\ }\href
  {https://doi.org/10.1063/1.4944722} {\bibfield  {journal} {\bibinfo
  {journal} {J. Math. Phys.}\ }\textbf {\bibinfo {volume} {57}},\ \bibinfo
  {pages} {032304} (\bibinfo {year} {2016})},\ \Eprint
  {https://arxiv.org/abs/1512.05630} {arXiv:1512.05630 [hep-ph]} \BibitemShut
  {NoStop}%
\bibitem [{\citenamefont {Remiddi}\ and\ \citenamefont
  {Tancredi}(2016)}]{Remiddi:2016gno}%
  \BibitemOpen
  \bibfield  {author} {\bibinfo {author} {\bibfnamefont {E.}~\bibnamefont
  {Remiddi}}\ and\ \bibinfo {author} {\bibfnamefont {L.}~\bibnamefont
  {Tancredi}},\ }\bibfield  {title} {\bibinfo {title} {{Differential equations
  and dispersion relations for Feynman amplitudes. The two-loop massive sunrise
  and the kite integral}},\ }\href
  {https://doi.org/10.1016/j.nuclphysb.2016.04.013} {\bibfield  {journal}
  {\bibinfo  {journal} {Nucl. Phys. B}\ }\textbf {\bibinfo {volume} {907}},\
  \bibinfo {pages} {400} (\bibinfo {year} {2016})},\ \Eprint
  {https://arxiv.org/abs/1602.01481} {arXiv:1602.01481 [hep-ph]} \BibitemShut
  {NoStop}%
\bibitem [{\citenamefont {Broedel}\ \emph
  {et~al.}(2018{\natexlab{c}})\citenamefont {Broedel}, \citenamefont {Duhr},
  \citenamefont {Dulat},\ and\ \citenamefont {Tancredi}}]{Broedel:2017siw}%
  \BibitemOpen
  \bibfield  {author} {\bibinfo {author} {\bibfnamefont {J.}~\bibnamefont
  {Broedel}}, \bibinfo {author} {\bibfnamefont {C.}~\bibnamefont {Duhr}},
  \bibinfo {author} {\bibfnamefont {F.}~\bibnamefont {Dulat}},\ and\ \bibinfo
  {author} {\bibfnamefont {L.}~\bibnamefont {Tancredi}},\ }\bibfield  {title}
  {\bibinfo {title} {{Elliptic polylogarithms and iterated integrals on
  elliptic curves II: an application to the sunrise integral}},\ }\href
  {https://doi.org/10.1103/PhysRevD.97.116009} {\bibfield  {journal} {\bibinfo
  {journal} {Phys. Rev. D}\ }\textbf {\bibinfo {volume} {97}},\ \bibinfo
  {pages} {116009} (\bibinfo {year} {2018}{\natexlab{c}})},\ \Eprint
  {https://arxiv.org/abs/1712.07095} {arXiv:1712.07095 [hep-ph]} \BibitemShut
  {NoStop}%
\bibitem [{\citenamefont {Verrill}(1996)}]{Verrill:1996}%
  \BibitemOpen
  \bibfield  {author} {\bibinfo {author} {\bibfnamefont {H.~A.}\ \bibnamefont
  {Verrill}},\ }\bibfield  {title} {\bibinfo {title} {{Root lattices and
  pencils of varieties}},\ }\href@noop {} {\bibfield  {journal} {\bibinfo
  {journal} {Journal of Mathematics of Kyoto University}\ }\textbf {\bibinfo
  {volume} {36}},\ \bibinfo {pages} {423} (\bibinfo {year} {1996})}\BibitemShut
  {NoStop}%
\bibitem [{\citenamefont {M{\"u}ller-Stach}\ \emph {et~al.}(2014)\citenamefont
  {M{\"u}ller-Stach}, \citenamefont {Weinzierl},\ and\ \citenamefont
  {Zayadeh}}]{Muller-Stach:2012tgj}%
  \BibitemOpen
  \bibfield  {author} {\bibinfo {author} {\bibfnamefont {S.}~\bibnamefont
  {M{\"u}ller-Stach}}, \bibinfo {author} {\bibfnamefont {S.}~\bibnamefont
  {Weinzierl}},\ and\ \bibinfo {author} {\bibfnamefont {R.}~\bibnamefont
  {Zayadeh}},\ }\bibfield  {title} {\bibinfo {title} {{Picard-Fuchs equations
  for Feynman integrals}},\ }\href {https://doi.org/10.1007/s00220-013-1838-3}
  {\bibfield  {journal} {\bibinfo  {journal} {Commun. Math. Phys.}\ }\textbf
  {\bibinfo {volume} {326}},\ \bibinfo {pages} {237} (\bibinfo {year}
  {2014})},\ \Eprint {https://arxiv.org/abs/1212.4389} {arXiv:1212.4389
  [hep-ph]} \BibitemShut {NoStop}%
\bibitem [{\citenamefont {P{\"o}gel}\ \emph {et~al.}(2022)\citenamefont
  {P{\"o}gel}, \citenamefont {Wang},\ and\ \citenamefont
  {Weinzierl}}]{Pogel:2022yat}%
  \BibitemOpen
  \bibfield  {author} {\bibinfo {author} {\bibfnamefont {S.}~\bibnamefont
  {P{\"o}gel}}, \bibinfo {author} {\bibfnamefont {X.}~\bibnamefont {Wang}},\
  and\ \bibinfo {author} {\bibfnamefont {S.}~\bibnamefont {Weinzierl}},\
  }\bibfield  {title} {\bibinfo {title} {{The three-loop equal-mass banana
  integral in {\ensuremath{\varepsilon}}-factorised form with meromorphic
  modular forms}},\ }\href {https://doi.org/10.1007/JHEP09(2022)062} {\bibfield
   {journal} {\bibinfo  {journal} {JHEP}\ }\textbf {\bibinfo {volume} {09}},\
  \bibinfo {pages} {062}},\ \Eprint {https://arxiv.org/abs/2207.12893}
  {arXiv:2207.12893 [hep-th]} \BibitemShut {NoStop}%
\bibitem [{\citenamefont {Broedel}\ \emph {et~al.}(2019)\citenamefont
  {Broedel}, \citenamefont {Duhr}, \citenamefont {Dulat}, \citenamefont
  {Marzucca}, \citenamefont {Penante},\ and\ \citenamefont
  {Tancredi}}]{Broedel:2019kmn}%
  \BibitemOpen
  \bibfield  {author} {\bibinfo {author} {\bibfnamefont {J.}~\bibnamefont
  {Broedel}}, \bibinfo {author} {\bibfnamefont {C.}~\bibnamefont {Duhr}},
  \bibinfo {author} {\bibfnamefont {F.}~\bibnamefont {Dulat}}, \bibinfo
  {author} {\bibfnamefont {R.}~\bibnamefont {Marzucca}}, \bibinfo {author}
  {\bibfnamefont {B.}~\bibnamefont {Penante}},\ and\ \bibinfo {author}
  {\bibfnamefont {L.}~\bibnamefont {Tancredi}},\ }\bibfield  {title} {\bibinfo
  {title} {{An analytic solution for the equal-mass banana graph}},\ }\href
  {https://doi.org/10.1007/JHEP09(2019)112} {\bibfield  {journal} {\bibinfo
  {journal} {JHEP}\ }\textbf {\bibinfo {volume} {09}},\ \bibinfo {pages}
  {112}},\ \Eprint {https://arxiv.org/abs/1907.03787} {arXiv:1907.03787
  [hep-th]} \BibitemShut {NoStop}%
\bibitem [{\citenamefont {B{\"o}nisch}\ \emph {et~al.}(2021)\citenamefont
  {B{\"o}nisch}, \citenamefont {Fischbach}, \citenamefont {Klemm},
  \citenamefont {Nega},\ and\ \citenamefont {Safari}}]{Bonisch:2020qmm}%
  \BibitemOpen
  \bibfield  {author} {\bibinfo {author} {\bibfnamefont {K.}~\bibnamefont
  {B{\"o}nisch}}, \bibinfo {author} {\bibfnamefont {F.}~\bibnamefont
  {Fischbach}}, \bibinfo {author} {\bibfnamefont {A.}~\bibnamefont {Klemm}},
  \bibinfo {author} {\bibfnamefont {C.}~\bibnamefont {Nega}},\ and\ \bibinfo
  {author} {\bibfnamefont {R.}~\bibnamefont {Safari}},\ }\bibfield  {title}
  {\bibinfo {title} {{Analytic structure of all loop banana integrals}},\
  }\href {https://doi.org/10.1007/JHEP05(2021)066} {\bibfield  {journal}
  {\bibinfo  {journal} {JHEP}\ }\textbf {\bibinfo {volume} {05}},\ \bibinfo
  {pages} {066}},\ \Eprint {https://arxiv.org/abs/2008.10574} {arXiv:2008.10574
  [hep-th]} \BibitemShut {NoStop}%
\bibitem [{\citenamefont {Broedel}\ \emph {et~al.}(2022)\citenamefont
  {Broedel}, \citenamefont {Duhr},\ and\ \citenamefont
  {Matthes}}]{Broedel:2021zij}%
  \BibitemOpen
  \bibfield  {author} {\bibinfo {author} {\bibfnamefont {J.}~\bibnamefont
  {Broedel}}, \bibinfo {author} {\bibfnamefont {C.}~\bibnamefont {Duhr}},\ and\
  \bibinfo {author} {\bibfnamefont {N.}~\bibnamefont {Matthes}},\ }\bibfield
  {title} {\bibinfo {title} {{Meromorphic modular forms and the three-loop
  equal-mass banana integral}},\ }\href
  {https://doi.org/10.1007/JHEP02(2022)184} {\bibfield  {journal} {\bibinfo
  {journal} {JHEP}\ }\textbf {\bibinfo {volume} {02}},\ \bibinfo {pages}
  {184}},\ \Eprint {https://arxiv.org/abs/2109.15251} {arXiv:2109.15251
  [hep-th]} \BibitemShut {NoStop}%
\bibitem [{\citenamefont {P{\"o}gel}\ \emph {et~al.}(2023)\citenamefont
  {P{\"o}gel}, \citenamefont {Wang},\ and\ \citenamefont
  {Weinzierl}}]{Pogel:2022vat}%
  \BibitemOpen
  \bibfield  {author} {\bibinfo {author} {\bibfnamefont {S.}~\bibnamefont
  {P{\"o}gel}}, \bibinfo {author} {\bibfnamefont {X.}~\bibnamefont {Wang}},\
  and\ \bibinfo {author} {\bibfnamefont {S.}~\bibnamefont {Weinzierl}},\
  }\bibfield  {title} {\bibinfo {title} {{Bananas of equal mass: any loop, any
  order in the dimensional regularisation parameter}},\ }\href
  {https://doi.org/10.1007/JHEP04(2023)117} {\bibfield  {journal} {\bibinfo
  {journal} {JHEP}\ }\textbf {\bibinfo {volume} {04}},\ \bibinfo {pages}
  {117}},\ \Eprint {https://arxiv.org/abs/2212.08908} {arXiv:2212.08908
  [hep-th]} \BibitemShut {NoStop}%
\bibitem [{\citenamefont {Duhr}(2025)}]{Duhr:2025tdf}%
  \BibitemOpen
  \bibfield  {author} {\bibinfo {author} {\bibfnamefont {C.}~\bibnamefont
  {Duhr}},\ }\bibfield  {title} {\bibinfo {title} {{Modular forms for
  three-loop banana integrals}},\ }\href
  {https://doi.org/10.1007/JHEP08(2025)218} {\bibfield  {journal} {\bibinfo
  {journal} {JHEP}\ }\textbf {\bibinfo {volume} {08}},\ \bibinfo {pages}
  {218}},\ \Eprint {https://arxiv.org/abs/2502.15325} {arXiv:2502.15325
  [hep-th]} \BibitemShut {NoStop}%
\bibitem [{\citenamefont {Liu}\ and\ \citenamefont {Ma}(2023)}]{Liu:2022chg}%
  \BibitemOpen
  \bibfield  {author} {\bibinfo {author} {\bibfnamefont {X.}~\bibnamefont
  {Liu}}\ and\ \bibinfo {author} {\bibfnamefont {Y.-Q.}\ \bibnamefont {Ma}},\
  }\bibfield  {title} {\bibinfo {title} {{AMFlow: A Mathematica package for
  Feynman integrals computation via auxiliary mass flow}},\ }\href
  {https://doi.org/10.1016/j.cpc.2022.108565} {\bibfield  {journal} {\bibinfo
  {journal} {Comput. Phys. Commun.}\ }\textbf {\bibinfo {volume} {283}},\
  \bibinfo {pages} {108565} (\bibinfo {year} {2023})},\ \Eprint
  {https://arxiv.org/abs/2201.11669} {arXiv:2201.11669 [hep-ph]} \BibitemShut
  {NoStop}%
\bibitem [{\citenamefont {Duhr}\ and\ \citenamefont
  {Dulat}(2019)}]{Duhr:2019tlz}%
  \BibitemOpen
  \bibfield  {author} {\bibinfo {author} {\bibfnamefont {C.}~\bibnamefont
  {Duhr}}\ and\ \bibinfo {author} {\bibfnamefont {F.}~\bibnamefont {Dulat}},\
  }\bibfield  {title} {\bibinfo {title} {{PolyLogTools \textemdash{} polylogs
  for the masses}},\ }\href {https://doi.org/10.1007/JHEP08(2019)135}
  {\bibfield  {journal} {\bibinfo  {journal} {JHEP}\ }\textbf {\bibinfo
  {volume} {08}},\ \bibinfo {pages} {135}},\ \Eprint
  {https://arxiv.org/abs/1904.07279} {arXiv:1904.07279 [hep-th]} \BibitemShut
  {NoStop}%
\bibitem [{\citenamefont {Naterop}\ \emph {et~al.}(2020)\citenamefont
  {Naterop}, \citenamefont {Signer},\ and\ \citenamefont
  {Ulrich}}]{Naterop:2019bzt}%
  \BibitemOpen
  \bibfield  {author} {\bibinfo {author} {\bibfnamefont {L.}~\bibnamefont
  {Naterop}}, \bibinfo {author} {\bibfnamefont {A.}~\bibnamefont {Signer}},\
  and\ \bibinfo {author} {\bibfnamefont {Y.}~\bibnamefont {Ulrich}},\
  }\bibfield  {title} {\bibinfo {title} {handyg---rapid numerical evaluation of
  generalised polylogarithms in fortran},\ }\href
  {https://doi.org/10.1016/j.cpc.2020.107165} {\bibfield  {journal} {\bibinfo
  {journal} {Comput. Phys. Commun.}\ }\textbf {\bibinfo {volume} {253}},\
  \bibinfo {pages} {107165} (\bibinfo {year} {2020})},\ \Eprint
  {https://arxiv.org/abs/1909.01656} {arXiv:1909.01656 [hep-ph]} \BibitemShut
  {NoStop}%
\bibitem [{\citenamefont {Bednyakov}\ \emph
  {et~al.}(2013{\natexlab{a}})\citenamefont {Bednyakov}, \citenamefont
  {Pikelner},\ and\ \citenamefont {Velizhanin}}]{Bednyakov:2012rb}%
  \BibitemOpen
  \bibfield  {author} {\bibinfo {author} {\bibfnamefont {A.~V.}\ \bibnamefont
  {Bednyakov}}, \bibinfo {author} {\bibfnamefont {A.~F.}\ \bibnamefont
  {Pikelner}},\ and\ \bibinfo {author} {\bibfnamefont {V.~N.}\ \bibnamefont
  {Velizhanin}},\ }\bibfield  {title} {\bibinfo {title} {{Anomalous dimensions
  of gauge fields and gauge coupling beta-functions in the Standard Model at
  three loops}},\ }\href {https://doi.org/10.1007/JHEP01(2013)017} {\bibfield
  {journal} {\bibinfo  {journal} {JHEP}\ }\textbf {\bibinfo {volume} {01}},\
  \bibinfo {pages} {017}},\ \Eprint {https://arxiv.org/abs/1210.6873}
  {arXiv:1210.6873 [hep-ph]} \BibitemShut {NoStop}%
\bibitem [{\citenamefont {Jegerlehner}\ \emph {et~al.}(2002)\citenamefont
  {Jegerlehner}, \citenamefont {Kalmykov},\ and\ \citenamefont
  {Veretin}}]{Jegerlehner:2001fb}%
  \BibitemOpen
  \bibfield  {author} {\bibinfo {author} {\bibfnamefont {F.}~\bibnamefont
  {Jegerlehner}}, \bibinfo {author} {\bibfnamefont {M.~Y.}\ \bibnamefont
  {Kalmykov}},\ and\ \bibinfo {author} {\bibfnamefont {O.}~\bibnamefont
  {Veretin}},\ }\bibfield  {title} {\bibinfo {title} {{MS versus pole masses of
  gauge bosons: Electroweak bosonic two loop corrections}},\ }\href
  {https://doi.org/10.1016/S0550-3213(02)00613-2} {\bibfield  {journal}
  {\bibinfo  {journal} {Nucl. Phys. B}\ }\textbf {\bibinfo {volume} {641}},\
  \bibinfo {pages} {285} (\bibinfo {year} {2002})},\ \Eprint
  {https://arxiv.org/abs/hep-ph/0105304} {arXiv:hep-ph/0105304} \BibitemShut
  {NoStop}%
\bibitem [{\citenamefont {Jegerlehner}\ \emph
  {et~al.}(2003{\natexlab{a}})\citenamefont {Jegerlehner}, \citenamefont
  {Kalmykov},\ and\ \citenamefont {Veretin}}]{Jegerlehner:2002em}%
  \BibitemOpen
  \bibfield  {author} {\bibinfo {author} {\bibfnamefont {F.}~\bibnamefont
  {Jegerlehner}}, \bibinfo {author} {\bibfnamefont {M.~Y.}\ \bibnamefont
  {Kalmykov}},\ and\ \bibinfo {author} {\bibfnamefont {O.}~\bibnamefont
  {Veretin}},\ }\bibfield  {title} {\bibinfo {title} {{MS-bar versus pole
  masses of gauge bosons. 2. Two loop electroweak fermion corrections}},\
  }\href {https://doi.org/10.1016/S0550-3213(03)00177-9} {\bibfield  {journal}
  {\bibinfo  {journal} {Nucl. Phys. B}\ }\textbf {\bibinfo {volume} {658}},\
  \bibinfo {pages} {49} (\bibinfo {year} {2003}{\natexlab{a}})},\ \Eprint
  {https://arxiv.org/abs/hep-ph/0212319} {arXiv:hep-ph/0212319} \BibitemShut
  {NoStop}%
\bibitem [{\citenamefont {Jegerlehner}\ \emph
  {et~al.}(2003{\natexlab{b}})\citenamefont {Jegerlehner}, \citenamefont
  {Kalmykov},\ and\ \citenamefont {Veretin}}]{Jegerlehner:2002er}%
  \BibitemOpen
  \bibfield  {author} {\bibinfo {author} {\bibfnamefont {F.}~\bibnamefont
  {Jegerlehner}}, \bibinfo {author} {\bibfnamefont {M.~Y.}\ \bibnamefont
  {Kalmykov}},\ and\ \bibinfo {author} {\bibfnamefont {O.}~\bibnamefont
  {Veretin}},\ }\bibfield  {title} {\bibinfo {title} {{Full two loop
  electroweak corrections to the pole masses of gauge bosons}},\ }\href
  {https://doi.org/10.1016/S0920-5632(03)80204-9} {\bibfield  {journal}
  {\bibinfo  {journal} {Nucl. Phys. B Proc. Suppl.}\ }\textbf {\bibinfo
  {volume} {116}},\ \bibinfo {pages} {382} (\bibinfo {year}
  {2003}{\natexlab{b}})},\ \Eprint {https://arxiv.org/abs/hep-ph/0212003}
  {arXiv:hep-ph/0212003} \BibitemShut {NoStop}%
\bibitem [{\citenamefont {Chetyrkin}\ and\ \citenamefont
  {Zoller}(2012)}]{Chetyrkin:2012rz}%
  \BibitemOpen
  \bibfield  {author} {\bibinfo {author} {\bibfnamefont {K.~G.}\ \bibnamefont
  {Chetyrkin}}\ and\ \bibinfo {author} {\bibfnamefont {M.~F.}\ \bibnamefont
  {Zoller}},\ }\bibfield  {title} {\bibinfo {title} {{Three-loop
  {\textbackslash}beta-functions for top-Yukawa and the Higgs self-interaction
  in the Standard Model}},\ }\href {https://doi.org/10.1007/JHEP06(2012)033}
  {\bibfield  {journal} {\bibinfo  {journal} {JHEP}\ }\textbf {\bibinfo
  {volume} {06}},\ \bibinfo {pages} {033}},\ \Eprint
  {https://arxiv.org/abs/1205.2892} {arXiv:1205.2892 [hep-ph]} \BibitemShut
  {NoStop}%
\bibitem [{\citenamefont {Bednyakov}\ \emph
  {et~al.}(2013{\natexlab{b}})\citenamefont {Bednyakov}, \citenamefont
  {Pikelner},\ and\ \citenamefont {Velizhanin}}]{Bednyakov:2012en}%
  \BibitemOpen
  \bibfield  {author} {\bibinfo {author} {\bibfnamefont {A.~V.}\ \bibnamefont
  {Bednyakov}}, \bibinfo {author} {\bibfnamefont {A.~F.}\ \bibnamefont
  {Pikelner}},\ and\ \bibinfo {author} {\bibfnamefont {V.~N.}\ \bibnamefont
  {Velizhanin}},\ }\bibfield  {title} {\bibinfo {title} {{Yukawa coupling
  beta-functions in the Standard Model at three loops}},\ }\href
  {https://doi.org/10.1016/j.physletb.2013.04.038} {\bibfield  {journal}
  {\bibinfo  {journal} {Phys. Lett. B}\ }\textbf {\bibinfo {volume} {722}},\
  \bibinfo {pages} {336} (\bibinfo {year} {2013}{\natexlab{b}})},\ \Eprint
  {https://arxiv.org/abs/1212.6829} {arXiv:1212.6829 [hep-ph]} \BibitemShut
  {NoStop}%
\bibitem [{\citenamefont {Bednyakov}\ \emph
  {et~al.}(2013{\natexlab{c}})\citenamefont {Bednyakov}, \citenamefont
  {Pikelner},\ and\ \citenamefont {Velizhanin}}]{Bednyakov:2013eba}%
  \BibitemOpen
  \bibfield  {author} {\bibinfo {author} {\bibfnamefont {A.~V.}\ \bibnamefont
  {Bednyakov}}, \bibinfo {author} {\bibfnamefont {A.~F.}\ \bibnamefont
  {Pikelner}},\ and\ \bibinfo {author} {\bibfnamefont {V.~N.}\ \bibnamefont
  {Velizhanin}},\ }\bibfield  {title} {\bibinfo {title} {{Higgs self-coupling
  beta-function in the Standard Model at three loops}},\ }\href
  {https://doi.org/10.1016/j.nuclphysb.2013.07.015} {\bibfield  {journal}
  {\bibinfo  {journal} {Nucl. Phys. B}\ }\textbf {\bibinfo {volume} {875}},\
  \bibinfo {pages} {552} (\bibinfo {year} {2013}{\natexlab{c}})},\ \Eprint
  {https://arxiv.org/abs/1303.4364} {arXiv:1303.4364 [hep-ph]} \BibitemShut
  {NoStop}%
\bibitem [{\citenamefont {Chen}\ \emph {et~al.}(2023)\citenamefont {Chen},
  \citenamefont {Li}, \citenamefont {Wang},\ and\ \citenamefont
  {Wang}}]{Chen:2022wit}%
  \BibitemOpen
  \bibfield  {author} {\bibinfo {author} {\bibfnamefont {L.-B.}\ \bibnamefont
  {Chen}}, \bibinfo {author} {\bibfnamefont {H.~T.}\ \bibnamefont {Li}},
  \bibinfo {author} {\bibfnamefont {J.}~\bibnamefont {Wang}},\ and\ \bibinfo
  {author} {\bibfnamefont {Y.}~\bibnamefont {Wang}},\ }\bibfield  {title}
  {\bibinfo {title} {{Analytic result for the top-quark width at
  next-to-next-to-leading order in QCD}},\ }\href
  {https://doi.org/10.1103/PhysRevD.108.054003} {\bibfield  {journal} {\bibinfo
   {journal} {Phys. Rev. D}\ }\textbf {\bibinfo {volume} {108}},\ \bibinfo
  {pages} {054003} (\bibinfo {year} {2023})},\ \Eprint
  {https://arxiv.org/abs/2212.06341} {arXiv:2212.06341 [hep-ph]} \BibitemShut
  {NoStop}%
\bibitem [{\citenamefont {Chen}\ \emph
  {et~al.}(2026{\natexlab{b}})\citenamefont {Chen}, \citenamefont {Chen},
  \citenamefont {Guan},\ and\ \citenamefont {Ma}}]{Chen:2023osm}%
  \BibitemOpen
  \bibfield  {author} {\bibinfo {author} {\bibfnamefont {L.}~\bibnamefont
  {Chen}}, \bibinfo {author} {\bibfnamefont {X.}~\bibnamefont {Chen}}, \bibinfo
  {author} {\bibfnamefont {X.}~\bibnamefont {Guan}},\ and\ \bibinfo {author}
  {\bibfnamefont {Y.-Q.}\ \bibnamefont {Ma}},\ }\bibfield  {title} {\bibinfo
  {title} {{Top-quark decay at next-to-next-to-next-to-leading order in QCD}},\
  }\href {https://doi.org/10.1103/dvlc-58hx} {\bibfield  {journal} {\bibinfo
  {journal} {Phys. Rev. D}\ }\textbf {\bibinfo {volume} {114}},\ \bibinfo
  {pages} {034005} (\bibinfo {year} {2026}{\natexlab{b}})},\ \Eprint
  {https://arxiv.org/abs/2309.01937} {arXiv:2309.01937 [hep-ph]} \BibitemShut
  {NoStop}%
\bibitem [{\citenamefont {Herren}\ and\ \citenamefont
  {Steinhauser}(2018)}]{Herren:2017dho}%
  \BibitemOpen
  \bibfield  {author} {\bibinfo {author} {\bibfnamefont {F.}~\bibnamefont
  {Herren}}\ and\ \bibinfo {author} {\bibfnamefont {M.}~\bibnamefont
  {Steinhauser}},\ }\bibfield  {title} {\bibinfo {title} {{Version 3 of RunDec
  and CRunDec}},\ }\href {https://doi.org/10.1016/j.cpc.2017.11.014} {\bibfield
   {journal} {\bibinfo  {journal} {Comput. Phys. Commun.}\ }\textbf {\bibinfo
  {volume} {224}},\ \bibinfo {pages} {333} (\bibinfo {year} {2018})},\ \Eprint
  {https://arxiv.org/abs/1703.03751} {arXiv:1703.03751 [hep-ph]} \BibitemShut
  {NoStop}%
\bibitem [{\citenamefont {Jegerlehner}\ \emph
  {et~al.}(2013{\natexlab{a}})\citenamefont {Jegerlehner}, \citenamefont
  {Kalmykov},\ and\ \citenamefont {Kniehl}}]{Jegerlehner:2012kn}%
  \BibitemOpen
  \bibfield  {author} {\bibinfo {author} {\bibfnamefont {F.}~\bibnamefont
  {Jegerlehner}}, \bibinfo {author} {\bibfnamefont {M.~Y.}\ \bibnamefont
  {Kalmykov}},\ and\ \bibinfo {author} {\bibfnamefont {B.~A.}\ \bibnamefont
  {Kniehl}},\ }\bibfield  {title} {\bibinfo {title} {{On the difference between
  the pole and the $\overline {MS}$ masses of the top quark at the electroweak
  scale}},\ }\href {https://doi.org/10.1016/j.physletb.2013.04.012} {\bibfield
  {journal} {\bibinfo  {journal} {Phys. Lett. B}\ }\textbf {\bibinfo {volume}
  {722}},\ \bibinfo {pages} {123} (\bibinfo {year} {2013}{\natexlab{a}})},\
  \Eprint {https://arxiv.org/abs/1212.4319} {arXiv:1212.4319 [hep-ph]}
  \BibitemShut {NoStop}%
\bibitem [{\citenamefont {Marquard}\ \emph {et~al.}(2016)\citenamefont
  {Marquard}, \citenamefont {Smirnov}, \citenamefont {Smirnov}, \citenamefont
  {Steinhauser},\ and\ \citenamefont {Wellmann}}]{Marquard:2016dcn}%
  \BibitemOpen
  \bibfield  {author} {\bibinfo {author} {\bibfnamefont {P.}~\bibnamefont
  {Marquard}}, \bibinfo {author} {\bibfnamefont {A.~V.}\ \bibnamefont
  {Smirnov}}, \bibinfo {author} {\bibfnamefont {V.~A.}\ \bibnamefont
  {Smirnov}}, \bibinfo {author} {\bibfnamefont {M.}~\bibnamefont
  {Steinhauser}},\ and\ \bibinfo {author} {\bibfnamefont {D.}~\bibnamefont
  {Wellmann}},\ }\bibfield  {title} {\bibinfo {title} {{$\overline{\rm
  MS}$-on-shell quark mass relation up to four loops in QCD and a general
  SU$(N)$ gauge group}},\ }\href {https://doi.org/10.1103/PhysRevD.94.074025}
  {\bibfield  {journal} {\bibinfo  {journal} {Phys. Rev. D}\ }\textbf {\bibinfo
  {volume} {94}},\ \bibinfo {pages} {074025} (\bibinfo {year} {2016})},\
  \Eprint {https://arxiv.org/abs/1606.06754} {arXiv:1606.06754 [hep-ph]}
  \BibitemShut {NoStop}%
\bibitem [{\citenamefont {Hempfling}\ and\ \citenamefont
  {Kniehl}(1995)}]{PhysRevD.51.1386}%
  \BibitemOpen
  \bibfield  {author} {\bibinfo {author} {\bibfnamefont {R.}~\bibnamefont
  {Hempfling}}\ and\ \bibinfo {author} {\bibfnamefont {B.~A.}\ \bibnamefont
  {Kniehl}},\ }\bibfield  {title} {\bibinfo {title} {Relation between the
  fermion pole mass and ms\ifmmode\bar\else\textasciimacron\fi{} yukawa
  coupling in the standard model},\ }\href
  {https://doi.org/10.1103/PhysRevD.51.1386} {\bibfield  {journal} {\bibinfo
  {journal} {Phys. Rev. D}\ }\textbf {\bibinfo {volume} {51}},\ \bibinfo
  {pages} {1386} (\bibinfo {year} {1995})}\BibitemShut {NoStop}%
\bibitem [{\citenamefont {Jegerlehner}\ \emph
  {et~al.}(2013{\natexlab{b}})\citenamefont {Jegerlehner}, \citenamefont
  {Kalmykov},\ and\ \citenamefont {Kniehl}}]{Jegerlehner:2013dpa}%
  \BibitemOpen
  \bibfield  {author} {\bibinfo {author} {\bibfnamefont {F.}~\bibnamefont
  {Jegerlehner}}, \bibinfo {author} {\bibfnamefont {M.~Y.}\ \bibnamefont
  {Kalmykov}},\ and\ \bibinfo {author} {\bibfnamefont {B.~A.}\ \bibnamefont
  {Kniehl}},\ }\bibfield  {title} {\bibinfo {title} {{About the EW contribution
  to the relation between pole and MS-masses of the top-quark in the Standard
  Model}},\ }\href {https://doi.org/10.22323/1.191.0190} {\bibfield  {journal}
  {\bibinfo  {journal} {PoS}\ }\textbf {\bibinfo {volume} {DIS2013}},\ \bibinfo
  {pages} {190} (\bibinfo {year} {2013}{\natexlab{b}})},\ \Eprint
  {https://arxiv.org/abs/1307.4226} {arXiv:1307.4226 [hep-ph]} \BibitemShut
  {NoStop}%
\bibitem [{\citenamefont {Kniehl}\ \emph {et~al.}(2015)\citenamefont {Kniehl},
  \citenamefont {Pikelner},\ and\ \citenamefont {Veretin}}]{Kniehl:2015nwa}%
  \BibitemOpen
  \bibfield  {author} {\bibinfo {author} {\bibfnamefont {B.~A.}\ \bibnamefont
  {Kniehl}}, \bibinfo {author} {\bibfnamefont {A.~F.}\ \bibnamefont
  {Pikelner}},\ and\ \bibinfo {author} {\bibfnamefont {O.~L.}\ \bibnamefont
  {Veretin}},\ }\bibfield  {title} {\bibinfo {title} {{Two-loop electroweak
  threshold corrections in the Standard Model}},\ }\href
  {https://doi.org/10.1016/j.nuclphysb.2015.04.010} {\bibfield  {journal}
  {\bibinfo  {journal} {Nucl. Phys. B}\ }\textbf {\bibinfo {volume} {896}},\
  \bibinfo {pages} {19} (\bibinfo {year} {2015})},\ \Eprint
  {https://arxiv.org/abs/1503.02138} {arXiv:1503.02138 [hep-ph]} \BibitemShut
  {NoStop}%
\bibitem [{\citenamefont {Kataev}\ and\ \citenamefont
  {Molokoedov}(2022)}]{Kataev:2022dua}%
  \BibitemOpen
  \bibfield  {author} {\bibinfo {author} {\bibfnamefont {A.~L.}\ \bibnamefont
  {Kataev}}\ and\ \bibinfo {author} {\bibfnamefont {V.~S.}\ \bibnamefont
  {Molokoedov}},\ }\bibfield  {title} {\bibinfo {title} {{Notes on Interplay
  between the QCD and EW Perturbative Corrections to the
  Pole-Running-to-Top-Quark Mass Ratio}},\ }\href
  {https://doi.org/10.1134/S0021364022600902} {\bibfield  {journal} {\bibinfo
  {journal} {JETP Lett.}\ }\textbf {\bibinfo {volume} {115}},\ \bibinfo {pages}
  {704} (\bibinfo {year} {2022})},\ \Eprint {https://arxiv.org/abs/2201.12073}
  {arXiv:2201.12073 [hep-ph]} \BibitemShut {NoStop}%
\end{thebibliography}%

\end{document}